\documentclass[10pt,a4paper,twocolumn]{article}

\usepackage[hmargin=16mm,top=20mm,bottom=22mm,columnsep=7mm]{geometry}
\usepackage[T1]{fontenc}
\usepackage[utf8]{inputenc}
\usepackage{amsmath,amsthm,mathtools}
\usepackage{libertinus}
\usepackage{libertinust1math}
\usepackage{bbold}
\usepackage{bm}
\usepackage{microtype}
\usepackage{booktabs}
\usepackage{graphicx}
\usepackage{subcaption}
\usepackage{xcolor}
\usepackage{titlesec}
\usepackage{enumitem}
\usepackage[most]{tcolorbox}
\usepackage{placeins}
\usepackage{flafter}
\usepackage{cite}
\usepackage{hyperref}
\usepackage{doi}
\usepackage[nameinlink,capitalise]{cleveref}
\titleformat{\section}{\large\bfseries}{\thesection}{0.7em}{}
\titleformat{\subsection}{\normalsize\bfseries}{\thesubsection}{0.7em}{}
\titleformat{\subsubsection}{\normalsize\itshape}{\thesubsubsection}{0.7em}{}
\titlespacing*{\section}{0pt}{2.6ex plus 0.6ex minus 0.3ex}{1.2ex plus 0.2ex}
\titlespacing*{\subsection}{0pt}{2.2ex plus 0.5ex minus 0.2ex}{0.9ex plus 0.2ex}
\titlespacing*{\subsubsection}{0pt}{1.8ex plus 0.4ex minus 0.2ex}{0.7ex plus 0.2ex}

\setlist[itemize]{leftmargin=1.2em,itemsep=2pt,topsep=4pt}

\hypersetup{
  pdftitle={Structure-factor witnesses of thermal entanglement: exact optimal covariance criteria for the XY chain},
  pdfauthor={Julia Mathe, Ayaka Usui, Otfried Guehne, and Giuseppe Vitagliano},
  colorlinks=true,
  linkcolor=blue!55!black,
  citecolor=blue!55!black,
  urlcolor=blue!55!black
}

\newcommand{\Tr}{\operatorname{Tr}}
\newcommand{\SEP}{\operatorname{SEP}}
\newcommand{\id}{\mathbb{1}}
\newcommand{\dd}{\mathrm{d}}
\newcommand{\e}{\mathrm{e}}
\newcommand{\ii}{\mathrm{i}}
\newcommand{\ket}[1]{\lvert #1\rangle}
\newcommand{\bra}[1]{\langle #1\rvert}
\newcommand{\braket}[2]{\langle #1\vert #2\rangle}
\newcommand{\ketbra}[1]{\lvert #1\rangle\!\langle #1\rvert}
\newcommand{\mean}[1]{\langle #1\rangle}
\newcommand{\Var}[1]{(\Delta #1)^2}
\newcommand{\cmc}{\mathrm{CMC}}
\newcommand{\bsa}{\mathrm{BSA}}
\newcommand{\sfac}{\mathcal{S}}
\newcommand{\T}{\mathsf{T}}

\theoremstyle{plain}
\newtheorem{theorem}{Theorem}
\newtheorem{proposition}{Proposition}
\newtheorem{corollary}{Corollary}
\theoremstyle{definition}
\newtheorem{observation}{Observation}
\crefname{observation}{Observation}{Observations}

\tcbset{thmbox/.style={breakable, enhanced jigsaw, colback=black!4, colframe=black!35,
  boxrule=0.5pt, arc=3pt, left=6pt, right=6pt, top=5pt, bottom=5pt,
  before skip=10pt plus 2pt, after skip=10pt plus 2pt}}
\tcolorboxenvironment{theorem}{thmbox}
\tcolorboxenvironment{proposition}{thmbox}
\tcolorboxenvironment{corollary}{thmbox}
\tcolorboxenvironment{observation}{thmbox}

\begin{document}

\twocolumn[{%
\begin{center}
{\LARGE\bfseries Structure-factor witnesses of thermal entanglement:\\[3pt]
exact optimal covariance criteria for the XY chain\par}
\vspace{14pt}
{\large Julia Math\'e\textsuperscript{1}, Ayaka Usui\textsuperscript{2}, Otfried G\"uhne\textsuperscript{3}, and Giuseppe Vitagliano\textsuperscript{1,4}\par}
\vspace{8pt}
{\small
\textsuperscript{1}Vienna Center for Quantum Science and Technology, Atominstitut, TU Wien, 1020 Vienna, Austria\\
\textsuperscript{2}Departament de F\'{i}sica, Universitat Aut\`onoma de Barcelona, 08193 Bellaterra, Spain\\
\textsuperscript{3}Naturwissenschaftlich-Technische Fakult\"at, Universit\"at Siegen, Walter-Flex-Stra\ss e 3, 57068 Siegen, Germany\\
\textsuperscript{4}Instituto de F\'isica Corpuscular, Universitat de Val\`encia -- Consejo Superior de Investigaciones Cient\'ificas,\\ Parc Cient\'ific, E-46980 Paterna, Valencia, Spain\par}
\vspace{14pt}
\begin{minipage}{0.86\textwidth}
\small
Which entanglement witness is optimal for a thermal many-body state, and what does its structure reveal about the equilibrium physics? We address these questions for the spin-$1/2$ XY chain in a transverse field, using the covariance-matrix criterion (CMC), which requires only one- and two-spin correlations.
We show that for thermal states the optimal CMC witness is fixed by the minima over momentum of the three static spin structure factors, which are accessible to scattering experiments and can be evaluated directly in the thermodynamic limit.
The optimal witnesses select the collective spin modes with the smallest thermal fluctuations, with their form dictated by symmetry rather than by criticality. They certify entanglement, and can also bound entanglement measures, up to temperatures of the order of the exchange coupling, including in states without any pairwise entanglement.
A systematic investigation of entanglement in small chains reveals that the CMC is nearly as powerful as a moment-matrix hierarchy supplied with all spin correlations, and close to the XX point it detects thermal states that are PPT across every bipartition. In addition, we observe that the same witnesses
can also certify entanglement generated by a sudden quench from a thermal state in which none is detected.
\end{minipage}
\end{center}
\vspace{18pt}
}]

\section{Introduction}
\label{sec:introduction}

Entanglement, introduced as a foundational concept of quantum theory, has become also an organizing principle of many-body physics, with relevance in characterizing phases of matter and also in the classical simulation of quantum dynamics~\cite{HorodeckiRMP2009,Amico2008,Laflorencie2016}. Despite its significance, certifying it in realistic quantum many-body systems is notoriously difficult, especially so in settings where the state is not pure but mixed, and
only a restricted set of observables can be obtained.
Entanglement certification and quantification in many-body states, especially at nonzero temperature, must therefore be based on few-body data, and it must be adapted to the correlations that the system actually develops: a generic criterion easily misses them. To address this challenge, it is useful to consider entanglement witnesses~\cite{HorodeckiRMP2009,GuehneToth2009,FrerotFadelLewenstein2023}, i.e., observables $W$ whose expectation value is negative for at least one entangled state,
and nonnegative for all separable states.

Many such witnesses have been proposed for spin models. They are based on the energy~\cite{DowlingDohertyBartlett2004,Toth2005}, on magnetic susceptibilities~\cite{WiesniakVedralBrukner2005,BruknerVedralZeilinger2006}, on collective-spin fluctuations and spin squeezing~\cite{Toth2004,TothKnappGuhneBriegel2007,TothKnappGuhneBriegel2009}, on static structure factors~\cite{KrammerEtAl2009,CramerPlenioWunderlich2011,CramerEtAl2013,MartyEtAl2014,MartyCramerPlenio2016} and on dynamic susceptibilities and the quantum Fisher information~\cite{HaukeEtAl2016,GabbrielliSmerziPezze2018,FrerotRoscilde2019}. Recently, this approach has even been applied to quantum magnets analyzed with neutron-scattering data~\cite{Mathew2020,ScheieEtAl2021,Laurell2021,LaurellEtAl2025}. While all of these examples constitute relevant witnesses, they are not necessarily optimal and thus likely to miss relevant entangled states. However, finding such optimal witnesses is even more difficult as it generally requires optimizing over the set of all witnesses. One way to deal with this problem is to use hierarchies of relaxations that approximate the set of entanglement witnesses with increasingly strict constraints, which can also be adapted to the input data accessible on the quantum state~\cite{DohertyParriloSpedalieri2004,NavascuesOwariPlenio2009PRL,BohnetWaldraffBraunGiraud2017,NavascuesBaccariAcin2021,FrerotBaccariAcin2022,TavakoliEtAl2024}. Hierarchies of semidefinite relaxations of this kind are a standard tool also beyond entanglement detection: they originate from polynomial optimization and the classical moment problem~\cite{Lasserre2001,Parrilo2003, PhysRevLett.127.060504}, they characterize the set of quantum correlations~\cite{NavascuesPironioAcin2007,NavascuesPironioAcin2008} or quantify multipartite entanglement \cite{weinbrenner2026complete}, and they provide certified bounds on ground-state energies and other ground-state properties of many-body systems~\cite{Mazziotti2004,BaumgratzPlenio2012,BarthelHubener2012,KullEtAl2024,WangEtAl2024}, as well as on the equilibrium properties of local Hamiltonians at nonzero temperature~\cite{FawziFawziScalet2024}.

In fact, the separability problem can be relaxed to semidefinite programs (SDPs) and hierarchies thereof, for instance those based on the partial transposition~\cite{Peres1996,HorodeckiEtAl1996,JungnitschMoroderGuhne2011}, on symmetric extensions~\cite{DohertyParriloSpedalieri2002,DohertyParriloSpedalieri2004}, or on moment matrices, which ask whether the measured correlations admit a separable model at all~\cite{BohnetWaldraffBraunGiraud2017,FrerotRoscilde2021, FrerotBaccariAcin2022}. Such relaxations return the optimal witness\footnote{Here and in the following, we call a witness optimal for given data if, among all witnesses of a given family with a fixed normalization, e.g., the witnesses dual to a given relaxation, it attains the smallest expectation value on these data, i.e., it is the most strongly violated one. This notion should not be confused with the optimality of entanglement witnesses in the sense of Ref.~\cite{LewensteinKrausCiracHorodecki2000}, which refers to witnesses whose set of detected states is maximal.} for the given data as the dual solution of the SDP, but the size of the SDP grows rapidly with the number of particles and with the order of the relaxation, so that they quickly become intractable. Where they can be solved, such numerically optimized witnesses are powerful and provide a very systematic way to search for optimal witnesses, but their output is usually in the form of large coefficient matrices that are difficult to understand and translate into practical measurement protocols. This raises the questions that motivate the present work: can we (i) efficiently and systematically search for optimal witnesses, (ii) understand them analytically, and (iii) connect their structure to the equilibrium physics of the underlying model?

We study these questions for the one-dimensional XY chain in a transverse field, a paradigmatic model of quantum statistical mechanics~\cite{LiebSchultzMattis1961, Pfeuty1970,BarouchMcCoyDresden1970,BarouchMcCoy1971,Sachdev2011}.  The model maps to free fermions, it contains both the Ising and the XX universality classes, and its correlation functions are known in closed form.  Its ground-state and thermal entanglement have been studied extensively through concurrence~\cite{ArnesenBoseVedral2001,Wang2001,OsborneNielsen2002,OsterlohEtAl2002}, entanglement entropies~\cite{VidalLatorreRicoKitaev2003,CalabreseCardy2004}, the geometric measure~\cite{WeiEtAl2005,Orus2008}, and entanglement witnesses, in particular those based on the energy~\cite{DowlingDohertyBartlett2004,Toth2005,GuhneTothBriegel2005,GuhneToth2006,IgloiToth2023}, which can also detect genuine multipartite entanglement~\cite{GuhneTothBriegel2005,GiampaoloHiesmayr2013,HofmannOsterlohGuhne2014}. See for example Refs.~\cite{Amico2008,Laflorencie2016} for reviews.

In this model it is possible to calculate two-spin correlators at arbitrary distances, as well as multi-spin correlators and hence the reduced states of a few spins~\cite{HofmannOsterlohGuhne2014}, which in principle allows one to use this marginal information in a systematic search for entanglement witnesses. In particular, to circumvent the computational complexity of broad SDP relaxations, we
employ the so-called covariance-matrix criterion (CMC), which can be written as a structured SDP search over a restricted family of entanglement witnesses, namely weighted sums of variances of arbitrary linear combinations of local observables~\cite{HofmannTakeuchi2003,Guhne2004,GuhneMechlerTothAdam2006,GuhneHyllusGittsovichEisert2007,GittsovichEtAl2008,GittsovichGuhne2010,GittsovichHyllusGuhne2010,LiuEtAl2024}.

\begin{figure*}[t]
\centering
\includegraphics[width=\linewidth]{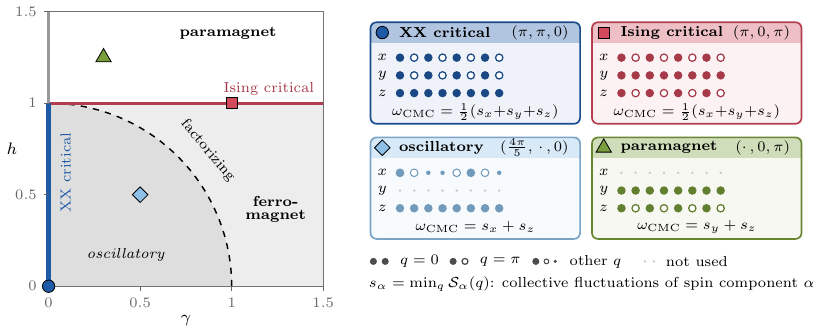}
\caption{Ground-state regimes of the XY chain \cref{eq:xy-hamiltonian} for $\gamma, h \geq 0$, with the coupling set to one (left), and the optimal covariance witness at four reference points (right). The Ising critical field $h = 1$ separates the ferromagnet from the paramagnet for every $\gamma > 0$. The segment $\gamma = 0$, $h < 1$ is the critical XX line. On the factorizing circle $h^2 + \gamma^2 = 1$ (dashed) the ground state is a product state, and inside it the ground-state correlations oscillate in space. Neither the circle nor the oscillatory region is a phase transition. For each spin component $\alpha$, the colored boxes show the signs $\e^{\ii q_\alpha j}$ with which consecutive sites enter the collective spin $J_\alpha(q_\alpha)$ whose fluctuations are smallest, as selected by \cref{thm:reduction} for the periodic chain with $N = 10$ at $T = 0.1$. Marker colors correspond to the witness types of \cref{fig:witness-map}(b).}
\label{fig:summary}
\end{figure*}

Our main results can be summarized as follows.
\begin{itemize}
\item \textit{Exact optimal witness.} Our central result is an exact solution for the optimal CMC witness of this model, spanning the entire coupling space and, crucially, nonzero temperatures. In the thermodynamic limit, it certifies entanglement up to temperatures of the order of the exchange coupling, about $2.5$ and $3.7$ times higher than the energy witness at the XX and Ising points (\cref{fig:entanglement-temperature}).
\item \textit{Physical content of the witness.} At the XX and Ising points the optimal witnesses are based on simple structure factors, i.e., staggered versions (right panel of \cref{fig:summary}) of the generalized spin-squeezing inequality $\sum_\alpha\Var{J_\alpha}\geq N/2$ for the collective spin $\bm J$~\cite{Toth2004,TothKnappGuhneBriegel2007}, which belongs to the family of spin-squeezing inequalities routinely used to detect entanglement in atomic ensembles~\cite{SorensenEtAl2001,KorbiczCiracLewenstein2005,VitaglianoEtAl2011,PezzeEtAl2018}. In general, only the minima over momentum of the three static structure factors are needed, which are accessible by neutron scattering in quantum magnets~\cite{CramerPlenioWunderlich2011,Mathew2020,ScheieEtAl2021,Laurell2021} and from measured spin correlations in quantum simulators with single-site resolution~\cite{GrossBloch2017,MonroeEtAl2021,BrowaeysLahaye2020}.
\item \textit{Benchmark in small chains.} Near the XX and Ising critical regions, the CMC is almost as strong as the more general moment-matrix hierarchy supplied with all Pauli correlations, and both detect entanglement in thermal states without any pairwise entanglement. Close to the XX point, the CMC even detects thermal states with a positive partial transpose (PPT) across every bipartition, for open as well as periodic chains of different lengths, as known also for spin-squeezing inequalities in thermal states of spin models~\cite{TothKnappGuhneBriegel2007,TothKnappGuhneBriegel2009}.
\item \textit{Quantification and dynamics.} The witness provides lower bounds on the entangled weight in the best separable approximation (BSA)~\cite{LewensteinSanpera1998,KarnasLewenstein2001} and on the geometric measure of entanglement. After a sudden quench from a thermal state in which no entanglement is detected, the CMC certifies the generated entanglement, which, as in the thermal states near detection threshold, is not carried by any pair of spins.
\end{itemize}

The paper is organized as follows.  In \Cref{sec:model-methods} we introduce the model and the separability criteria used in the work.
In \Cref{sec:small-chain-results} we present the small-chain benchmark, in which the three criteria are compared for thermal states of four spins with complete information on the state.  \Cref{sec:analytic-witness} contains the main results. First, we describe the optimal covariance witnesses obtained numerically for longer chains in terms of static structure factors. Afterwards, we derive the closed form of the optimal covariance witness from the symmetries of the thermal state, and discuss its physical content and its thermodynamic limit, where it yields the highest temperature at which entanglement is detected. In \Cref{sec:quench} we apply the CMC to a quench from a thermal state which is initially not detected, and we observe that entanglement is generated globally, but not locally in the state. I.e., entanglement is detected via the CMC in the steady state, even though all two-body marginals are separable. Finally, in \Cref{sec:quantification} we discuss bounds on distance measures of entanglement, in particular on the entangled weight in the best separable approximation and on the geometric measure of entanglement. We then conclude in \cref{sec:conclusions}. Further technical details and numerical investigations are collected in the appendices.

\section{Preliminary tools}
\label{sec:model-methods}

In this section we introduce the model and the tools used in the rest of the paper. In \cref{subsec:xy-model} we define the thermal XY chain and discuss its symmetries, its exact solution and its equilibrium regimes, and we fix the two reference points of our analysis. In \cref{subsec:entanglement-framework} we define full separability and entanglement witnesses, and we introduce the three entanglement criteria that we compare in \cref{sec:small-chain-results}: the PPT criterion, the CMC and the moment-matrix hierarchy.

\subsection{The thermal XY chain}
\label{subsec:xy-model}

We consider $N$ qubits with Hamiltonian\footnote{Note that the sign of $\mathcal J$ is not relevant here as for OBC, and for PBC with even $N$, the staggered rotation $\sigma_j^{x,y}\mapsto(-1)^j\sigma_j^{x,y}$ maps $\mathcal J$ to $-\mathcal J$ and, for PBC, merely interchanges the momenta $q=0$ and $q=\pi$ of the transverse spin components.}
\begin{equation}
H=-\mathcal J\sum_{j=1}^{N_b}
\left[
\frac{1+\gamma}{2}\,\sigma_j^x\sigma_{j+1}^x
+\frac{1-\gamma}{2}\,\sigma_j^y\sigma_{j+1}^y
\right]
-h\sum_{j=1}^{N}\sigma_j^z ,
\label{eq:xy-hamiltonian}
\end{equation}
where $\mathcal J>0$ is the exchange coupling, $\gamma$ is the anisotropy and $h$ the transverse field.  For periodic boundary conditions (PBC) $N_b=N$ and $\sigma_{N+1}^\alpha= \sigma_1^\alpha$, while for open boundary conditions (OBC) $N_b=N-1$.  We consider the Gibbs state
\begin{equation}
\varrho_\beta=\frac{\e^{-\beta H}}{\Tr\e^{-\beta H}},
\label{eq:thermal-state}
\end{equation}
where $\beta = 1/T$, we set $k_{\rm B}=1$, and in what follows we measure the temperature and the external field in units of $\mathcal J$, i.e., we set $\mathcal J=1$ (and $\hbar=1$).

Since $\varrho_\beta$ is a function of $H$ alone, it automatically shares all symmetries of the Hamiltonian. For PBC, $H$ and thus $\varrho_\beta$ are invariant under lattice translations, so that all two-spin correlators $\mean{\sigma_i^\alpha\sigma_j^\beta}$ depend only on the distance between the two spins. For all $\gamma$ and $h$, $H$ commutes with the spin parity $P_z=\prod_j\sigma_j^z$, which rotates every spin by $\pi$ around the $z$ axis, $\sigma_j^{x,y}\mapsto-\sigma_j^{x,y}$. This forces $\mean{\sigma_j^x}=\mean{\sigma_j^y}=0$ and makes all correlations between a transverse and a longitudinal spin component vanish. Finally, $H$ is a real matrix in the $\sigma^z$ eigenbasis, so that $\varrho_\beta=\varrho_\beta^*$. This time-reversal-like symmetry acts on the local Bloch vectors as the reflection $\sigma^y\mapsto-\sigma^y$, and it forces the symmetrized correlations between $\sigma^x$ and $\sigma^y$ to vanish. These three symmetries together greatly simplify the optimal witness, as will be discussed in the following. On the XX line, $\gamma=0$, the spin parity is enlarged to the continuous $U(1)$ symmetry of simultaneous rotations of all spins about the $z$ axis. It is generated by the total magnetization $J_z=\frac12\sum_j\sigma_j^z$, which is conserved, $[H,J_z]=0$, and it makes the correlations of the $x$ and $y$ components identical.

As is well known, this model can be solved via the Jordan--Wigner transformation, which maps \cref{eq:xy-hamiltonian} to a quadratic fermionic Hamiltonian, which a Bogoliubov transformation brings into the form $H=\sum_k\varepsilon_k(d_k^\dagger d_k-\frac12)$ with the single-fermion dispersion~\cite{LiebSchultzMattis1961,Pfeuty1970}
\begin{equation}
\varepsilon_k=2\sqrt{\left(\cos k-h\right)^2+\gamma^2\sin^2 k} ,
\label{eq:dispersion}
\end{equation}
where $d_k^\dagger$ and $d_k$ are fermionic creation and annihilation operators of Bogoliubov quasiparticles with momentum $k$, $\{d_k,d_{k'}^\dagger\}=\delta_{kk'}$. From this mapping it is also possible to calculate the spin correlation functions, which follow from Wick's theorem: the longitudinal ones are products of fermionic two-point functions, the transverse ones can be obtained as Toeplitz determinants~\cite{BarouchMcCoyDresden1970,BarouchMcCoy1971}, i.e., for two spins at distance $r$, as determinants of $r\times r$ matrices of fermionic two-point functions. The computational effort thus grows only polynomially with $N$ and $r$, so that chains of hundreds of spins are accessible and, since the fermionic two-point functions are then given by integrals over momentum, so is the thermodynamic limit. The details
are given in \cref{app:xy-solution}.

For $\gamma\neq0$ the ground state is ordered for $|h|<1$: it spontaneously breaks the spin-parity symmetry and develops long-range ferromagnetic order of the $x$ components of the spins for $\gamma>0$ (of the $y$ components for $\gamma<0$), i.e., the correlations $\mean{\sigma_0^x\sigma_r^x}$ tend to a nonzero value $m_x^2$ at large distances $r$, where $m_x$ is the spontaneous magnetization~\cite{BarouchMcCoy1971}. For $|h|>1$, instead, the ground state is a paramagnet polarized along the field.  The two regimes are separated by Ising transitions at $|h|=1$, at which the low-energy physics is described by a conformal field theory with central charge $c=1/2$, that of a free Majorana fermion~\cite{BelavinPolyakovZamolodchikov1984,Cardy1996,DiFrancesco1997,CalabreseCardy2004,Mussardo2010,DuttaEtAl2015,MbengRussomannoSantoro2024}.  The segment $\gamma=0$, $|h|<1$ is instead the critical XX phase, with the low-energy physics described by a $c=1$ Luttinger liquid~\cite{Haldane1981,Voit1995,GogolinNersesyanTsvelik1998,Giamarchi2003,CazalillaEtAl2011}. The XX line thus separates the $x$- and $y$-ordered phases, with endpoints $(\gamma,h)=(0,\pm1)$. Inside the ordered region, the circle $h^2+\gamma^2=1$ hosts a factorized ground state~\cite{KurmannThomasMuller1982,MullerShrock1985,RoscildeEtAl2004,AmicoEtAl2006,Amico2008,GiampaoloAdessoIlluminati2008,RossignoliCanosaMatera2008,GiampaoloAdessoIlluminati2009}. More precise statements on these regimes, including the critical exponents, the definition of the correlation length and its behavior at nonzero temperature, are collected in \cref{app:critical}. In the light of this (see also the left panel of \cref{fig:summary}), we choose as our two reference points the center of the XX line, $(\gamma,h)=(0,0)$, and the point $(1,1)$ on the Ising critical line $h=1$ (which we call the Ising point).

\subsection{Separability criteria}
\label{subsec:entanglement-framework}

In this work, we investigate how to witness $N$-qubit states that are {\it entangled}, i.e., not fully separable. The latter are formally defined as follows: An $N$-qubit state is fully separable if it is a convex combination of products of single-qubit states,
\begin{equation}
\begin{gathered}
    \varrho_{\rm sep}=\sum_s p_s \left( \varrho_1 \otimes \dots \otimes \varrho_N \right)_s , \\
 \text{with} \quad \varrho_j = \frac{\id+\bm r_j^{(s)}\cdot\bm\sigma_j}{2}, \\
p_s\geq0,\quad \sum_s p_s=1,\quad
|\bm r_j^{(s)}|\leq1 .
\end{gathered}
\label{eq:full-separable-decomposition}
\end{equation}
Here $\bm\sigma_j=(\sigma_j^x,\sigma_j^y,\sigma_j^z)$ is the vector of Pauli matrices, and $\bm r_j^{(s)}$ is the (so-called ``Bloch'') vector of the spin at position $j$, in the product state labeled by $s$. The condition $|\bm r_j^{(s)}|\leq1$ (i.e., bounded Bloch-vector length) follows from the positivity of the single-spin density matrix $(\id+\bm r\cdot\bm\sigma)/2$,
with pure states satisfying $|\bm r|=1$ (i.e., being on the surface of the Bloch sphere). We denote the set of fully separable states by $\SEP$ and look for entanglement witnesses, i.e., operators that can certify that $\varrho\notin\SEP$. Formally, an entanglement witness is a Hermitian operator $W$ such that:
\begin{equation}
    \begin{aligned}
        \Tr(W\varrho_{\rm sep})&\geq0 \quad \text{for all} \ \varrho_{\rm sep}\in\SEP , \\
        \Tr(W\varrho)&<0 \quad \text{for at least one} \ \varrho \notin \SEP .
    \end{aligned}
\end{equation}
Hence, a negative expectation value of $W$ certifies entanglement~\cite{HorodeckiRMP2009,GuehneToth2009}. More generally, we also use nonlinear witnesses, i.e., functionals of quantum states that are not necessarily linear but are bounded from below on the set of separable states, such as the variance-based criteria introduced below.

We consider essentially three canonical methods that provide entanglement witnesses in generic states, as detailed below.

\subsubsection[The partial-transposition criterion]{The partial-transposition criterion}

The first is based on the partial transposition, i.e., on the so-called PPT criterion. A state separable with respect to a bipartition $A|\bar{A}$ has a positive partial transpose, $\varrho^{T_A}\succeq0$~\cite{Peres1996,HorodeckiEtAl1996}. Thus, a negative eigenvalue of the partially transposed state certifies entanglement, and its amount can also be quantified as
\begin{equation}
   \mathcal N_{A|\bar A}=
(\lVert\varrho^{T_A}\rVert_1-1)/2 ,
\end{equation}
which is called {\it entanglement negativity}~\cite{ZyczkowskiEtAl1998,VidalWerner2002,Plenio2005}.  A state that has a positive partial transpose across every bipartition, which we call {\it fully PPT}, need not be fully separable, as positivity under partial transposition is only a necessary condition for separability, and thus its violation only a sufficient condition for entanglement. In practice, the partial transpose is conveniently computed in the Pauli basis. Writing $\varrho=2^{-N}\sum_{\bm\mu}c_{\bm\mu}\,\sigma_1^{\mu_1}\otimes\cdots\otimes\sigma_N^{\mu_N}$ with $\mu_j\in\{0,x,y,z\}$, $\sigma^0=\id$, and real coefficients $c_{\bm\mu}=\mean{\sigma_1^{\mu_1}\otimes\cdots\otimes\sigma_N^{\mu_N}}_\varrho$, and using that the transposition leaves $\id$, $\sigma^x$ and $\sigma^z$ invariant and maps $\sigma^y$ to $(\sigma^y)^\T=-\sigma^y$, one obtains $\varrho^{T_A}$ from $\varrho$ by flipping the sign of all terms with an odd number of $\sigma^y$ factors on the spins in $A$. For a given cut, $W_A=(P_-)^{T_A}$, where $P_-$ projects onto the eigenvectors of $\varrho^{T_A}$ with negative eigenvalues, is a PPT witness~\cite{GuehneToth2009} with $v_{\rm PPT}=\Tr(W_A\varrho) =-\mathcal N_{A|\bar A}$, which is negative exactly for states that are not PPT across this cut. Indeed, $\Tr(W_A\varrho)=\Tr(P_-\varrho^{T_A})$ is the sum of the negative eigenvalues of $\varrho^{T_A}$, whereas $\Tr(W_A\varrho_{\rm sep})=\Tr(P_-\varrho_{\rm sep}^{T_A})\geq0$ for every state $\varrho_{\rm sep}$ that is separable across $A|\bar A$.

The PPT criterion is also the lowest level of complete hierarchies of SDP relaxations of the set of separable states, most notably the hierarchy of PPT symmetric extensions~\cite{DohertyParriloSpedalieri2002,DohertyParriloSpedalieri2004}, which applies also to multipartite entanglement~\cite{DohertyParriloSpedalieri2005}, and related complete criteria~\cite{NavascuesOwariPlenio2009PRL,NavascuesOwariPlenio2009PRA}. Such relaxations have been adapted to the marginals that are accessible in many-body systems~\cite{NavascuesBaccariAcin2021}, and PPT-based criteria have been applied extensively to XY and Ising chains, e.g., through the negativity between blocks of spins~\cite{WichterichEtAl2009,CalabreseTagliacozzoTonni2013} and through PPT mixtures~\cite{HofmannOsterlohGuhne2014}. Relaxations based on the PPT criterion also provide tests of genuine multipartite entanglement (GME), which we discuss in \cref{app:gme}.

\subsubsection[The covariance-matrix criterion]{The covariance-matrix criterion}

The second test, the CMC, uses only the first and second moments of the local Pauli operators, collected in a $3N\times3N$ {\it covariance matrix}:
\begin{equation}
\Gamma_{j\alpha,l\beta}(\varrho)
=\frac12\mean{\{\sigma_j^\alpha,\sigma_l^\beta\}}_{\varrho}
-\mean{\sigma_j^\alpha}_{\varrho}\mean{\sigma_l^\beta}_{\varrho} ,
\qquad \alpha,\beta\in\{x,y,z\} .
\label{eq:spin-covariance}
\end{equation}
Explicitly, $\Gamma$ consists of $3\times3$ blocks as
\begin{equation}
\begin{gathered}
\Gamma=\begin{pmatrix}
\Gamma_{11} & \Gamma_{12} & \cdots & \Gamma_{1N}\\
\Gamma_{21} & \Gamma_{22} & \cdots & \Gamma_{2N}\\
\vdots & \vdots & \ddots & \vdots\\
\Gamma_{N1} & \Gamma_{N2} & \cdots & \Gamma_{NN}
\end{pmatrix},
\\[4pt]
\Gamma_{jl}=\begin{pmatrix}
\Gamma_{jx,lx} & \Gamma_{jx,ly} & \Gamma_{jx,lz}\\
\Gamma_{jy,lx} & \Gamma_{jy,ly} & \Gamma_{jy,lz}\\
\Gamma_{jz,lx} & \Gamma_{jz,ly} & \Gamma_{jz,lz}
\end{pmatrix}.
\end{gathered}
\label{eq:cm-blocks}
\end{equation}
\noindent Since $(\sigma_j^\alpha)^2=\id$ and $\{\sigma_j^\alpha,\sigma_j^\beta\}=0$ for $\alpha\neq\beta$, the diagonal blocks $\Gamma_{jj}=\id_3-\bm r_j\bm r_j^\T$ depend only on the local Bloch vectors $\bm r_j=\mean{\bm\sigma_j}$, whereas the off-diagonal blocks contain the connected two-spin correlations $\Gamma_{j\alpha,l\beta}=\mean{\sigma_j^\alpha\sigma_l^\beta}-\mean{\sigma_j^\alpha}\mean{\sigma_l^\beta}$, $j\neq l$.
Ordering the rows and columns by spin component instead, $\Gamma$ consists of the $N\times N$ blocks $\Gamma^{\alpha\beta}$ with entries $(\Gamma^{\alpha\beta})_{jl}=\Gamma_{j\alpha,l\beta}$.
In particular, a single qubit in a pure state
has covariance matrix $\id_3-\bm r\bm r^\T$ with Bloch vector $\bm r$ satisfying $|\bm r|=1$.
Together with the concavity of the covariance matrix under mixing, this implies that every fully separable state satisfies~\cite{GuhneHyllusGittsovichEisert2007,GittsovichEtAl2008,GittsovichGuhne2010,GittsovichHyllusGuhne2010}
\begin{equation}
\Gamma(\varrho)\succeq\bigoplus_{j=1}^N(\id_3-X_j),
\qquad X_j\succeq0,\quad \Tr X_j=1 ,
\label{eq:cmc-primal}
\end{equation}
for some real symmetric $3\times3$ matrices $X_j$.  We call a covariance matrix of this form a \textit{separable covariance matrix}: it is the sum of the local noise $\bigoplus_j(\id_3-X_j)$ of pure product states and of a positive semidefinite remainder, which accounts for classical correlations. Thus, for a given $N$-qubit state $\varrho$ and its corresponding covariance matrix $\Gamma(\varrho)$, the primal problem is to find a separable covariance matrix that bounds it as in \cref{eq:cmc-primal}. The dual formulation provides the entanglement witnesses~\cite{Hyllus06,GittsovichEtAl2008,MatheBose2026}.  For every real $Z\succeq0$ with diagonal $3\times3$ blocks $Z_{jj}$, every fully separable state obeys
\begin{equation}
\Tr[Z\Gamma(\varrho)]\geq b(Z):=\sum_{j=1}^N
\left[\Tr Z_{jj}-\lambda_{\max}(Z_{jj})\right].
\label{eq:cmc-dual-bound}
\end{equation}
Optimizing over all $Z$ with a fixed normalization gives
\begin{equation}
\omega_{\cmc}(\varrho)=\min_{Z\succeq0,\ b(Z)\geq1}\Tr[Z\Gamma(\varrho)] ,
\label{eq:cmc-optimum}
\end{equation}
and $\omega_{\cmc}(\varrho)<1$ certifies entanglement.  By a separating-hyperplane argument~\cite{GittsovichEtAl2008,GittsovichGuhne2010,GittsovichHyllusGuhne2010}, $\omega_{\cmc}(\varrho)\geq1$ holds if and only if $\Gamma(\varrho)$ is a separable covariance matrix in the sense of \cref{eq:cmc-primal}, as in typical convex duality arguments of semidefinite programs.  Writing $Z=\sum_\nu z_\nu\bm v_\nu\bm v_\nu^\T$ shows that
\begin{equation}
\Tr[Z\Gamma(\varrho)]=\sum_\nu z_\nu\Var{A_\nu}_{\varrho},
\qquad \text{with} \quad
A_\nu=\sum_{j,\alpha}(\bm v_\nu)_{j\alpha}\,\sigma_j^\alpha ,
\label{eq:cmc-lur}
\end{equation}
so the CMC is an optimized local uncertainty relation~\cite{HofmannTakeuchi2003,GittsovichEtAl2008} for collective observables (in this case generalized collective spin components or structure factors, as we will clarify below).

\subsubsection[The moment-matrix hierarchy]{The moment-matrix hierarchy}

The third test, a moment-matrix hierarchy, exploits what can be called an {\it entanglement marginal problem}~\cite{BohnetWaldraffBraunGiraud2017,NavascuesBaccariAcin2021}, i.e., it asks whether the measured Pauli correlations can be reproduced by a classical distribution of local Bloch vectors~\cite{BohnetWaldraffBraunGiraud2017,FrerotBaccariAcin2022}.  If $\varrho$ is fully separable, there is a probability measure $\mu$ on the set of Bloch vectors $\bm x=(\bm r_1,\dots,\bm r_N)$, $|\bm r_j|\leq1$, such that
\begin{equation}
\mean{\sigma_{j_1}^{\alpha_1}\cdots\sigma_{j_m}^{\alpha_m}}_{\varrho}
=\int r_{j_1}^{\alpha_1}\cdots r_{j_m}^{\alpha_m}\,\dd\mu(\bm x)
\qquad (j_1,\dots,j_m\ \text{distinct}) ,
\label{eq:pauli-classical-moments}
\end{equation}
which can be then relaxed to a truncated sequence of moments, where the measured data appear as given elements of a moment matrix, together with unconstrained auxiliary moments (i.e., those not measured).

For instance, for $N=2$ spins and the lowest order, the relaxation asks whether there are real symmetric $3\times3$ matrices $X_1$ and $X_2$ with $\Tr X_1\leq1$ and $\Tr X_2\leq1$ such that
\begin{equation}
\begin{pmatrix}
1 & \bm m_1^\T & \bm m_2^\T\\
\bm m_1 & X_1 & C_{12}\\
\bm m_2 & C_{12}^\T & X_2
\end{pmatrix}\succeq0 ,
\label{eq:moment-matrix-example}
\end{equation}
where the means $(\bm m_j)_\alpha=\mean{\sigma_j^\alpha}_\varrho$ and the correlations $(C_{12})_{\alpha\beta}=\mean{\sigma_1^\alpha\sigma_2^\beta}_\varrho$ are measured. For a separable state, the matrix in \cref{eq:moment-matrix-example} is the matrix $\int\bm v\bm v^\T\dd\mu$ of the second moments of the vector $\bm v=(1,\bm r_1,\bm r_2)$ of all monomials of degree at most one, the auxiliary moments are $(X_j)_{\alpha\beta}=\int r_j^\alpha r_j^\beta\,\dd\mu$, which should satisfy $\Tr X_j=\int|\bm r_j|^2\,\dd\mu\leq1$. The auxiliary moments are not fixed by the data, since \cref{eq:pauli-classical-moments} holds only for distinct spins. For instance, $\int(r_1^x)^2\,\dd\mu$ is not related to $\mean{(\sigma_1^x)^2}_\varrho=1$. At higher orders, the moment matrix contains also moments of higher degree, e.g., three- and four-spin correlations, and the resulting hierarchy of relaxations is complete, i.e., all data that admit no separable model are detected at a sufficiently high order~\cite{BohnetWaldraffBraunGiraud2017,Nie2014}.

In general, for $N$ spins and relaxation order $k$, the moments of all monomials in the $3N$ Bloch-vector components $r_j^\alpha$ up to degree $2k$ form a sequence $y$, which is arranged in a moment matrix $M_k(y)$ whose rows and columns are labeled by the monomials of degree at most $k$. Through \cref{eq:pauli-classical-moments}, the data fix the moments of products of components of distinct spins, whereas all other moments are free variables. A separable model requires $M_k(y)\succeq0$, together with the positivity of so-called localizing matrices, which encode the constraints $|\bm r_j|\leq1$ and generalize the conditions $\Tr X_j\leq1$ of the example above~\cite{BohnetWaldraffBraunGiraud2017}. The explicit form of these semidefinite conditions is given in \cref{app:moment-matrix}.

As for the CMC, the infeasibility certifies entanglement, and the dual SDP returns a witness that is linear in the supplied Pauli moments. It corresponds to a polynomial in the Bloch-vector components of the form
\begin{equation}
\begin{split}
p(\bm x)={}&\bm m_k(\bm x)^\T Q_0\,\bm m_k(\bm x)\\
&+\sum_{j=1}^N\bigl(1-|\bm r_j|^2\bigr)\,\bm m_{k-1}(\bm x)^\T Q_j\,\bm m_{k-1}(\bm x),
\end{split}
\label{eq:sos-certificate}
\end{equation}
where $\bm m_k(\bm x)$ is the vector of all monomials of degree at most $k$ and $Q_0\succeq0$, $Q_j\succeq0$.
We denote by $v_{\rm SDP}$ its value on the data, normalized such that the matrices $Q_0$ and $Q_j$
have total trace at most one, $\Tr Q_0+\sum_j\Tr Q_j\leq1$, i.e., $v_{\rm SDP}$ is the minimum of $\sum_{\bm a}c_{\bm a}\mean{\sigma^{\bm a}}_\varrho$ over all certificates of the form \cref{eq:sos-certificate} with this normalization
and $v_{\rm SDP}<0$ certifies entanglement. Feasibility alone is inconclusive, since a feasible sequence $y$ need not be the moment sequence of any measure. However, eventually it is possible to verify that the solution is {\it flat}, i.e., $\operatorname{rank}M_k=\operatorname{rank} M_{k-1}$, and this implies that the measure can be explicitly extracted, thereby yielding an explicit separable state compatible with the data (see \cite{BohnetWaldraffBraunGiraud2017} and references therein).

Since infeasibility at any order excludes a separable model, a flat solution makes the hierarchy conclusive at a finite order, i.e., necessary and sufficient for the given data. Flatness is, however, not guaranteed at any finite order, and the
hierarchy is complete only in the limit $k\to\infty$~\cite{BohnetWaldraffBraunGiraud2017,Nie2014}. The problem nevertheless remains computationally demanding already for relatively small systems: the moment matrix $M_k$ has $\binom{3N+k}{k}$ rows, e.g., $91$ for $N=4$ and $k=2$ but $5456$ for $N=10$ and $k=3$, and the order needed for detection, or for flatness, is not known in advance. This reflects the fact that deciding separability is NP-hard in general~\cite{Gurvits2003,Gharibian2010}.
As we discuss in \cref{sec:quantification}, explicit separable decompositions of this kind are also what is needed to bound entanglement monotones from above.

If only one- and two-spin correlations are supplied, the first level of the hierarchy is equivalent to the CMC, since the free local second moments then play the role of the matrices $X_j$ in \cref{eq:cmc-primal} (\cref{app:moment-matrix}).
The moment-matrix hierarchy thus builds upon the CMC, which is its first level for one- and two-spin data, whereas the PPT criterion is the lowest level of the hierarchies of PPT symmetric extensions~\cite{DohertyParriloSpedalieri2002,DohertyParriloSpedalieri2004,DohertyParriloSpedalieri2005} and of related relaxations~\cite{NavascuesOwariPlenio2009PRL,NavascuesOwariPlenio2009PRA,NavascuesBaccariAcin2021} mentioned above. Here we are interested in hierarchies that build upon the CMC, since their lowest level requires only one- and two-spin correlations and remains tractable for long chains.

\section{Benchmark in small chains}
\label{sec:small-chain-results}

For $N=4$ all three tests can be run with complete information on the state. We construct $\varrho_\beta$ by exact diagonalization, evaluate the partial transpose across all bipartitions, solve the CMC with the complete local-Pauli covariance matrix, and run the moment hierarchy at relaxation order $k=2$ with all Pauli correlations as data.  In all our data the alternating cut $\{1,3\}|\{2,4\}$ gives the most negative partial transpose whenever any of the seven partial transposes is negative, and the figures show the PPT witness for this cut. This agrees with area-law arguments: in thermal states of local Hamiltonians, the total correlations between the two sides of a cut are bounded by a quantity proportional to the inverse temperature and to the number of interaction bonds that cross it, rather than to the size of the two parts~\cite{WolfEtAl2008,EisertCramerPlenio2010}, and the negativity obeys a similar area law~\cite{ShermanEtAl2016}. The alternating cut separates the two sublattices of the chain and is crossed by every nearest-neighbor bond, three for OBC and four for PBC, whereas every other cut crosses at most two.

\begin{figure*}[t]
    \centering
    \includegraphics[width=1.0\linewidth]{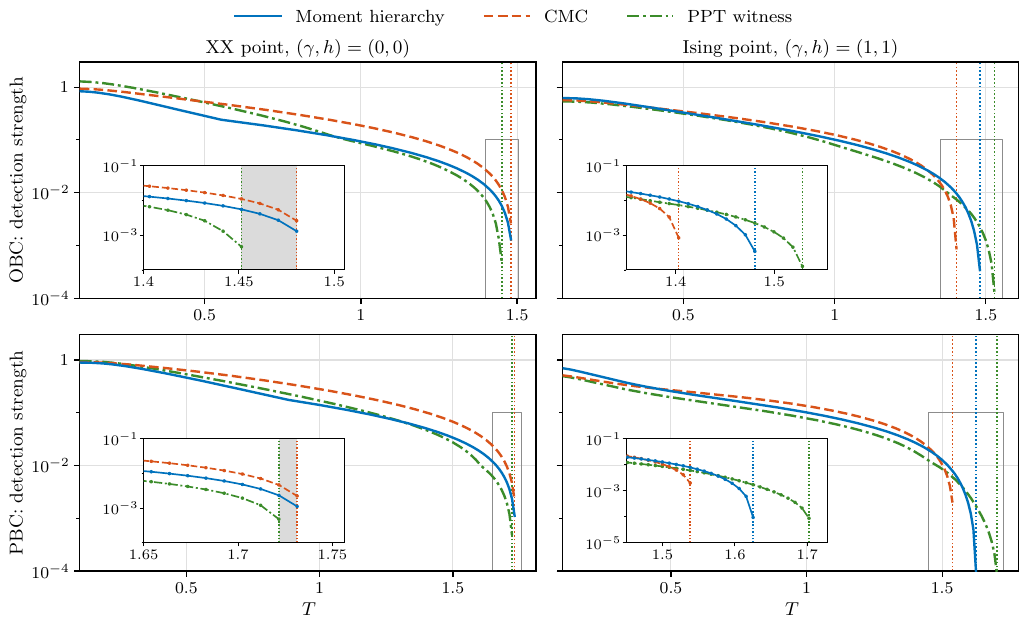}
    \caption{Detection strength of the criteria introduced in \cref{subsec:entanglement-framework} as a function of temperature for $N=4$ with open (OBC, top row) and periodic (PBC, bottom row) boundary conditions, at the XX point (left column) and at the Ising point (right column), with all three criteria in each panel on a logarithmic scale. Moment hierarchy: $\max(0,-v_{\rm SDP})$; CMC: $\max(0,1-\omega_{\cmc})$; PPT witness for the alternating cut: $\max(0,-v_{\rm PPT})$.
    Dotted vertical lines mark the highest temperature at which each criterion detects entanglement.
    The insets zoom in on these temperatures, in the windows marked by grey frames, with dots at the computed temperatures. At the XX point (left column), the CMC and the moment hierarchy detect entanglement up to higher temperatures than the PPT witness, and the shaded interval in the insets lies between the highest detection temperatures of the PPT witness and of the CMC. Three further points are shown in \cref{fig:criteria-comparison-app}.
    }
\label{fig:criteria-comparison}
\end{figure*}

In \cref{fig:criteria-comparison}, we compare the different criteria at the two reference points as a function of temperature, and in \cref{fig:criteria-comparison-app} (\cref{app:small-chain-maps}) at three further points of the phase diagram. For low temperatures, all criteria detect all states. For increasing temperatures, differences in detection strength become apparent, and the two boundary conditions lead to the same qualitative picture, with higher detection temperatures for PBC. At the XX point, $h=\gamma=0$, the moment hierarchy and the CMC detect entanglement up to the same temperature, $T\simeq1.49$ for OBC and $T\simeq1.74$ for PBC, slightly above the PPT test, which stops at $T\simeq1.46$ and $T\simeq1.73$, respectively. At the Ising point, and at the three further points of \cref{fig:criteria-comparison-app}, the order is reversed: the PPT test detects entanglement up to temperatures at least as high as the moment hierarchy, which in turn exceeds the CMC. Between these temperatures, shaded in the insets of the first column of \cref{fig:criteria-comparison}, the thermal state at the XX point is PPT across every bipartition, so that its entanglement is invisible to the PPT test but certified by the CMC, as found also for spin-squeezing inequalities in thermal states of spin models~\cite{TothKnappGuhneBriegel2007,TothKnappGuhneBriegel2009}, and such states also appear around the XX point in the $(\gamma,h)$ plane and in longer chains with both boundary conditions (\cref{app:small-chain-maps}).

The comparison between the CMC and the PPT test contains another conceptually important observation: the CMC detects entanglement in states without any pairwise entanglement. At both reference points and $T=1.47$, for instance, all two-spin reduced states of the periodic chain are PPT, and hence separable, while $\omega_{\cmc}=0.92$ at the XX point and $\omega_{\cmc}=0.98$ at the Ising point.  Pairwise measures such as the concurrence, which have been the standard tool in early studies of entanglement in spin chains~\cite{OsborneNielsen2002,OsterlohEtAl2002}, miss this collective entanglement.

Comparing the two moment-based criteria, the CMC is almost as strong as the moment hierarchy supplied with all Pauli correlations, and at the XX point equally strong. Complementary numerical tests that we present in \cref{app:small-chain-maps} show that
the hierarchy detects every state detected by the CMC and, in addition, states close to the boundary of the CMC detection region. This indicates that the additional detection power relies on three- and four-spin correlations.

Focusing on the PPT witness, it is also evident that there exists a region for larger $h$ and $\gamma$, illustrated by $h = 1.20, \gamma = 1.30$ (\cref{fig:criteria-comparison-app}), where this criterion detects more than both the hierarchy and the CMC. Since for $N=4$ the full set of Pauli correlations determines the state, higher levels of the hierarchy would eventually detect every state that is not PPT. The advantage of the PPT test in this region therefore reflects the finite relaxation order $k=2$ of the hierarchy, whereas the CMC is in addition restricted to one- and two-spin correlations. Finally, the PPT-mixture relaxation, which tests for genuine multipartite entanglement, certifies it in the same states only up to considerably lower temperatures (\cref{app:gme}).

To summarize, the moment hierarchy detects at least as much as the CMC and is the only test that can return an explicit separable decomposition, but its matrices grow rapidly with the relaxation order and with $N$. The PPT test needs the full density matrix and addresses one bipartition at a time. The CMC needs only a $3N\times3N$ covariance matrix, which is accessible from one- and two-spin measurements, and it is nearly as effective as the moment hierarchy precisely in the critical regions. In the rest of the paper we therefore focus on the CMC and ask what its optimal witness looks like.

\section{The optimal covariance witness in closed form}
\label{sec:analytic-witness}

\subsection{Numerical optimum in larger chains}
\label{subsec:cmc-critical-numerics}

\begin{figure*}[t]
    \centering
    \includegraphics[width=1.0\linewidth]{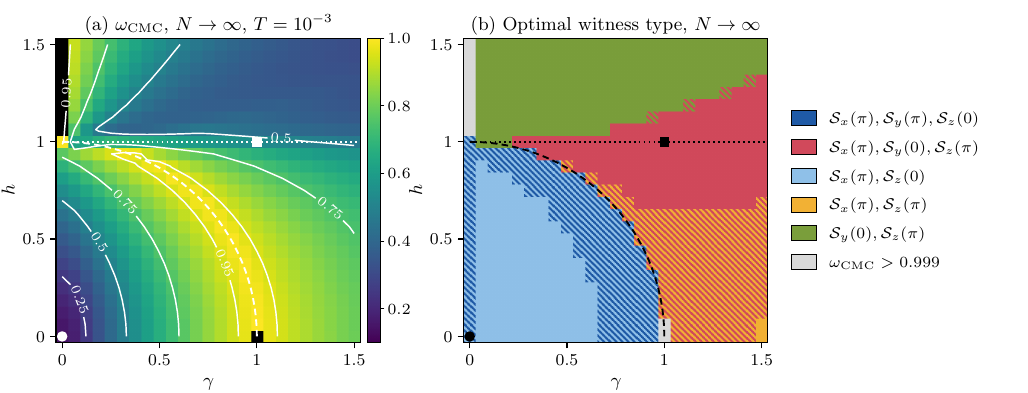}
    \caption{Optimal covariance witness in the $(\gamma,h)$ plane in the thermodynamic limit, $N\to\infty$, at $T=10^{-3}$, from \cref{eq:exact-cmc,eq:structure-factor-tdl}. (a)~Optimal value $\omega_{\cmc}$; entanglement is detected wherever $\omega_{\cmc}<1$, and black pixels are not detected.  (b)~Spin modes entering the optimal witness: three-mode witnesses of XX and Ising type and planar witnesses with two components. Hatched pixels mark points at which two of these witnesses differ by less than $1\%$. On the XX line, the planar witness built from $\sfac_y(\pi)$ and $\sfac_z(0)$ is equivalent to the one built from $\sfac_x(\pi)$ and $\sfac_z(0)$ by symmetry.  Grey points have $\omega_{\cmc}>0.999$, where the structure factors are nearly flat and the minimizing momenta are not meaningful.  The dashed line is the factorizing circle $h^2+\gamma^2=1$, the dotted line the Ising critical field $h=1$; the circle and the square mark the XX and Ising points.  In the ordered region the thermal correlation length exceeds the range of distances that we sum, and the structure factors are evaluated at the momenta $0$ and $\pi$ only (\cref{app:xy-solution}). Inside the circle, where the minimum of $\sfac_x$ can move to incommensurate momenta, panel (a) is therefore an upper bound on $\omega_{\cmc}$ and panel (b) shows the best witness with commensurate momenta.}
    \label{fig:witness-map}
\end{figure*}

Since the CMC is an SDP over $3N\times3N$ matrices, it can be solved for much longer chains than the other two tests.  For a window of $80$ spins in the middle of an open chain of $100$ spins at $T=10^{-3}$, for instance, it detects entanglement in essentially the entire $(\gamma,h)$ plane of \cref{fig:witness-map}, except on the XX line above the saturation field and at $(\gamma,h)=(1,0)$, where the Gibbs state of the classical Ising chain is separable; on the factorizing circle $h^2+\gamma^2=1$, which passes through this point, the violation is at best marginal, $\omega_{\cmc}\simeq1$ (see below). The same holds in the thermodynamic limit shown in \cref{fig:witness-map}(a), which is not obtained by extrapolation but directly from \cref{thm:reduction}, as described in \cref{subsec:thermodynamic-limit} below, i.e., as the limit of periodic chains.

\begin{figure*}[t]
\centering
\begin{subfigure}[b]{.85\textwidth}
\includegraphics[width=\linewidth]{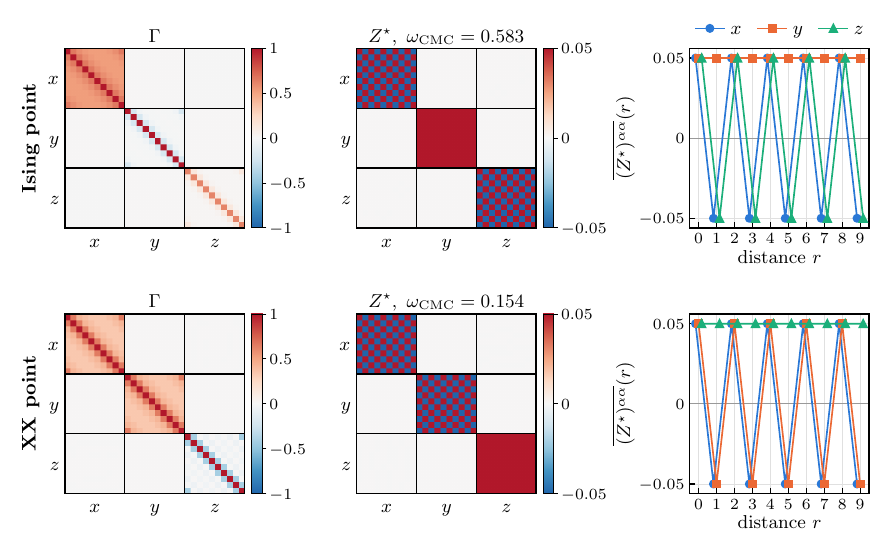}
\caption{Periodic chain}
\end{subfigure}\vfill
\begin{subfigure}[b]{.85\textwidth}
\includegraphics[width=\linewidth]{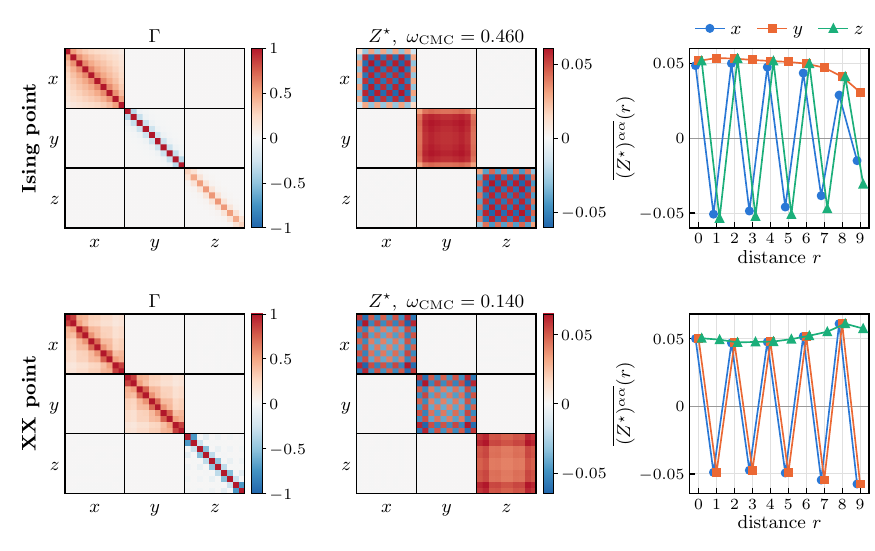}
\caption{Open chain}
\end{subfigure}
\caption{Covariance matrix $\Gamma$, optimal CMC matrix $Z^\star$, and entries of the spin blocks $(Z^\star)^{\alpha\alpha}$ averaged over pairs of sites at distance $r$ for the periodic (top) and open (bottom) chain with $N=10$ at $T=0.1$, at the Ising point (upper row of each panel) and at the XX point (lower row). The matrices are ordered by spin component, $x$, $y$, $z$. For PBC, $Z^\star$ is one half of a rank-three projector onto one Fourier mode in each spin sector: $(x,\pi)$, $(y,0)$, $(z,\pi)$ at the Ising point and $(x,\pi)$, $(y,\pi)$, $(z,0)$ at the XX point. \Cref{thm:reduction} explains this structure. For OBC, $Z^\star$ deviates from this form close to the edges. The optimal values $\omega_{\cmc}$ are given above the panels of $Z^\star$.
}
\label{fig:cmc-optimal-Z}
\end{figure*}

To look inside the optimizer we consider chains of $N=10$ spins at $T=0.1$. The witness values and optimal matrices $Z^\star$ are displayed in \cref{fig:cmc-optimal-Z}.
Especially for PBC, they have a strikingly simple structure: $Z^\star$ has exactly three nonzero eigenvalues, all equal to $1/2$,
and each eigenvector is a single Fourier mode of a single spin component.  At the XX point the three modes are $(x,\pi)$, $(y,\pi)$ and $(z,0)$, where $(\alpha,q)$ denotes spin component $\alpha$ at lattice momentum $q$. At the Ising point they are $(x,\pi)$, $(y,0)$ and $(z,\pi)$.  For OBC, and for windows of longer open chains, the optimal matrix is no longer exactly a Fourier projector. For windows in the bulk of long open chains, however, the optimal values are close to those of the ring of the same size and approach the thermodynamic limit as the window grows. At the Ising point, for example, a window of $10$ spins in the middle of an open chain of $200$ spins gives $\omega_{\cmc}=0.591$, compared to $0.583$ for the ring with $N=10$ and $0.559$ for $N\to\infty$. Open boundaries, a bulk subchain, and a noncritical point are discussed in \cref{app:small-chain-maps}.

\subsection[\texorpdfstring{Analytic structure of the CMC optimum}{Analytic structure of the CMC optimum}]{Analytic structure of the CMC optimum}

The structure of $Z^\star$ (cf.~\cref{fig:cmc-optimal-Z}) suggests describing the optimal witnesses through collective fluctuations at fixed momentum, namely to consider the connected static structure factors
\begin{equation}
\begin{gathered}
\sfac_\alpha(q)=\frac1N\sum_{j,l=1}^N\e^{\ii q(j-l)}\,\Gamma_{j\alpha,l\alpha}(\varrho),
\\
q\in\mathcal Q_N=\Bigl\{\tfrac{2\pi m}{N}:m=0,\dots,N-1\Bigr\},
\end{gathered}
\label{eq:structure-factor}
\end{equation}
and in particular their minima over momentum,
\begin{equation}
s_\alpha=\min_{q\in\mathcal Q_N}\sfac_\alpha(q),
\qquad \alpha\in\{x,y,z\},
\label{eq:structure-factor-minima}
\end{equation}
attained at momenta $q_\alpha$.  For every momentum $q$, the structure factor is a variance of a (in general non-Hermitian) collective spin component,
\begin{equation}
\sfac_\alpha(q)=\frac4N\Var{J_\alpha(q)},
\qquad
J_\alpha(q)=\frac12\sum_{j=1}^N\e^{\ii qj}\sigma_j^\alpha ,
\label{eq:momentum-spin}
\end{equation}
which can be used to derive entanglement criteria based on uncertainty relations, following well-established methods~\cite{HofmannTakeuchi2003,Guhne2004,Toth2004,GittsovichEtAl2008,TothKnappGuhneBriegel2009,KrammerEtAl2009,CramerPlenioWunderlich2011}. For $q\notin\{0,\pi\}$ the operator $J_\alpha(q)$ is not Hermitian, and we define the variance of such an operator $A$ as $\Var{A}_\varrho=\mean{A^\dagger A}_\varrho-\mean{A^\dagger}_\varrho\mean{A}_\varrho$. Since $[J_\alpha(q),J_\alpha(q)^\dagger]=0$, this is the sum of the variances of the commuting Hermitian operators $[J_\alpha(q)+J_\alpha(q)^\dagger]/2$ and $[J_\alpha(q)-J_\alpha(q)^\dagger]/(2\ii)$, i.e., of the cosine and sine components of $J_\alpha(q)$.

\begin{observation}[Collective uncertainty relations criteria]
\label{prop:three-mode}
For every $N$, every choice of real momenta $\bm q=(q_x,q_y,q_z)$, and every fully separable state $\varrho_{\rm sep}$,
\begin{equation}
\begin{gathered}
\omega_{\bm q}(\varrho_{\rm sep})\geq1 ,
\\
\text{where}\quad
\omega_{\bm q}(\varrho):=\frac2N\sum_{\alpha=x,y,z}\Var{J_\alpha(q_\alpha)}_{\varrho}
 .
\end{gathered}
\label{eq:three-mode-witness}
\end{equation}
\end{observation}

\begin{proof} For pure product states $\ket\psi=\bigotimes_j\ket{\psi_j}$
with unit Bloch vectors $\bm r_j$ we have $\mean{\sigma_j^\alpha}_\psi=r_j^\alpha$, $(\sigma_j^\alpha)^2=\id$ and $\mean{\sigma_j^\alpha\sigma_l^\alpha}_\psi=r_j^\alpha r_l^\alpha$ for $j\neq l$. Hence only the terms with $j=l$ contribute to the variances,
\begin{equation}
\begin{split}
\Var{J_\alpha(q_\alpha)}_\psi&=\frac14\sum_{j,l=1}^N\e^{\ii q_\alpha(j-l)}
\bigl(\mean{\sigma_j^\alpha\sigma_l^\alpha}_\psi-r_j^\alpha r_l^\alpha\bigr)\\
&=\frac14\sum_{j=1}^N\bigl[1-(r_j^\alpha)^2\bigr] .
\end{split}
\end{equation}
Summing over the three spin components gives
\begin{equation}
\sum_{\alpha=x,y,z}\Var{J_\alpha(q_\alpha)}_\psi
=\frac14\sum_{j=1}^N\bigl(3-|\bm r_j|^2\bigr)=\frac N2 ,
\end{equation}
i.e., $\omega_{\bm q}(\psi)=1$. A fully separable state is a mixture $\varrho_{\rm sep}=\sum_sp_s\ketbra{\psi_s}$ of pure product states, and the variance is concave, $\Var{A}_{\varrho_{\rm sep}}\geq\sum_sp_s\Var{A}_{\psi_s}$, also for non-Hermitian $A$, since $\mean{A^\dagger A}$ is linear and $|\mean{A}|^2$ is convex in the state. Hence $\omega_{\bm q}(\varrho_{\rm sep})\geq\sum_sp_s\,\omega_{\bm q}(\psi_s)=1$.
\end{proof}

\begin{table*}[t]
\centering
\caption{Minimal structure factors $s_\alpha$, minimizing momenta $q_\alpha$ and optimal CMC value at $T=0.1$ for the periodic chain with $N=10$ and in the thermodynamic limit ($N=\infty$).  In all cases
$\omega_{\cmc}=\frac12(s_x+s_y+s_z)$.}
\label{tab:structure-factor-minima}
\begin{tabular}{@{}lcccccc@{}}
\toprule
 & $N$ & $s_x$ & $s_y$ & $s_z$ & $(q_x,q_y,q_z)$ & $\omega_{\cmc}$ \\
\midrule
XX point $(0,0)$ & $10$ & $0.138$ & $0.138$ & $0.031$ & $(\pi,\pi,0)$ & $0.154$ \\
 & $\infty$ & $0.157$ & $0.157$ & $0.064$ & $(\pi,\pi,0)$ & $0.189$ \\
Ising point $(1,1)$ & $10$ & $0.288$ & $0.485$ & $0.393$ & $(\pi,0,\pi)$ & $0.583$ \\
 & $\infty$ & $0.299$ & $0.453$ & $0.366$ & $(\pi,0,\pi)$ & $0.559$ \\
\bottomrule
\end{tabular}
\end{table*}

\begin{figure*}[t]
\centering
\includegraphics[width=\textwidth]{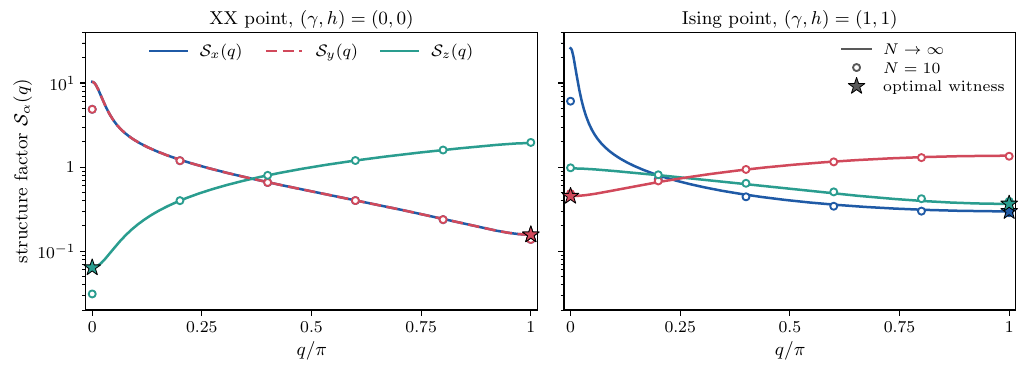}
\caption{Connected static structure factors $\sfac_\alpha(q)$ at $T=0.1$ at the XX point (left) and at the Ising point (right).  Lines: thermodynamic limit, from the free-fermion solution (\cref{app:xy-solution}); circles: periodic chain with $N=10$ (exact diagonalization).  At the XX point $\sfac_x=\sfac_y$.  Stars mark the minima selected by the optimal witness, $(q_x,q_y,q_z)=(\pi,\pi,0)$ and $(\pi,0,\pi)$, respectively.
}
\label{fig:structure-factors}
\end{figure*}

In particular, at our two reference points the optimal CMC witnesses can be identified as
\begin{align}
\omega_{\rm XX}(\varrho)&=\frac2N\Bigl[\Var{J_x(\pi)}+\Var{J_y(\pi)}+\Var{J_z(0)}\Bigr]_{\varrho} ,
\label{eq:xx-witness}\\
\omega_{\rm Ising}(\varrho)&=\frac2N\Bigl[\Var{J_x(\pi)}+\Var{J_y(0)}+\Var{J_z(\pi)}\Bigr]_{\varrho} ,
\label{eq:ising-witness}
\end{align}
which means that entanglement is certified by three structure factors at the momenta $0$ and $\pi$. The optimal witnesses thus combine, in each spin component, the collective fluctuations of \cref{eq:momentum-spin} that are smallest, which, as we also observe in the proof of \cref{thm:reduction}, is a consequence of the symmetries of the thermal state.

Away from the two reference points, the minimizing momenta, and even the number of spin components that enter the optimal witness, change in a way that reflects the equilibrium structure of the model.  \Cref{fig:witness-map} shows $\omega_{\cmc}$ and the type of optimal witness for $0\leq\gamma\leq1.5$ and $0\leq h\leq1.5$, in the thermodynamic limit at $T=10^{-3}$, obtained from \cref{eq:exact-cmc}. On the XX line the $U(1)$ symmetry fixes $s_x=s_y$, and the XX witness in \cref{eq:xx-witness} is optimal.  Since $s_z=T\chi$, see \cref{eq:xx-susceptibility}, is tiny at this temperature, the planar witnesses introduced below, which use only one of the two transverse components, are almost as good (hatched pixels).  A small anisotropy $\gamma>0$ enhances the $y$ fluctuations, and as soon as $s_y>s_x+s_z$ the optimal witness drops the $y$ component altogether.  It then becomes a two-component, or planar, uncertainty relation~\cite{HofmannTakeuchi2003,VitaglianoEtAl2018,HePengDrummondReid2011,DammeierSchwonnekWerner2015,SzymanskiZyczkowski2020},
\begin{equation}
\Var{J_x(\pi)}+\Var{J_z(0)}\geq\frac N4 ,
\label{eq:planar-witness}
\end{equation}
whose normalized violation is $s_x+s_z$.  For $N=10$, $T=0.1$ and $h=0$ this happens already at $\gamma\simeq0.013$, and at $\gamma=0.05$ the violation is even slightly stronger than on the XX line, $\omega_{\cmc}=0.144$.  In large parts of the ordered region, however, $s_y$ remains close to $s_x+s_z$, so that the three-mode and the planar witnesses differ by less than $1\%$ (hatched regions). This is the case in a band inside the circle $h^2+\gamma^2=1$, for the XX witness and \cref{eq:planar-witness}, and outside the circle for $h\lesssim0.6$, for the Ising witness and the planar witness built from $J_x(\pi)$ and $J_z(\pi)$.  In a broad region around the Ising critical line and outside the circle, the Ising witness \cref{eq:ising-witness} is optimal.  In the field-polarized phase, where the $z$ fluctuations are small, the optimum is the planar witness built from $J_y(0)$ and $J_z(\pi)$.

Inside the factorizing circle, the spin correlations oscillate in space with a period that changes continuously with $\gamma$ and $h$ and is in general incommensurate with the lattice~\cite{BarouchMcCoy1971}. Accordingly, the minimum of $\sfac_x(q)$ can move away from the commensurate momenta $q=0$ and $q=\pi$, which correspond to uniform and staggered spin patterns, so that the least fluctuating collective $x$ mode is a spin modulation with a period longer than two sites, as we find for $N=10$ (see~\cref{app:small-chain-maps}). In the ordered region at $T=10^{-3}$, however, we evaluate the structure factors only at $q=0$ and $q=\pi$ (\cref{app:xy-solution}), so that \cref{fig:witness-map}(a) shows the value of the best witness built from these two commensurate modes. The circle itself, on which the ground state factorizes, is where the witness is weakest, $\omega_{\cmc}\simeq1$, and the same holds on the XX line above the saturation field, where the ground state is fully polarized.  Across the circle, the minimizing momentum of $\sfac_z$ jumps from $q_z=0$ to $q_z=\pi$, and that of $\sfac_y$ from $q_y=\pi$ to $q_y=0$, consistent with a change of sign of the connected longitudinal and $y$ correlations, which vanish in the factorized ground state.  The optimal witness thus reflects the factorizing circle, the incommensurate region and, more loosely, the Ising critical line, without any input beyond the covariance matrix. See also \cref{app:small-chain-maps} for additional numerics for $N=10$.

The low rank and the clean momentum content of $Z^\star$ are a direct consequence of the symmetries of the thermal state. In particular, this allows us to find explicitly a general structure of the optimal inequalities of this type for the case of the XY chain, exploiting its structure and symmetries.

\begin{theorem}[Optimal CMC for the thermal XY chain]
\label{thm:reduction}
Let $\varrho$ be the Gibbs state \eqref{eq:thermal-state} of the XY chain \eqref{eq:xy-hamiltonian} with PBC, for arbitrary $N$, $\gamma$, $h$ and $T>0$. Then
\begin{equation}
\!\omega_{\cmc}(\varrho)=\min\Bigl\{\tfrac12(s_x\!+\!s_y\!+\!s_z),\;\!\!
s_x+s_y,\;\!\! s_y+s_z,\;\!\! s_z+s_x\!\Bigr\}.
\label{eq:exact-cmc}
\end{equation}
\end{theorem}

\vspace{0.5cm}

\noindent In other words, the optimal CMC witness is given by the matrix $Z^\star=\sum_\alpha\kappa_\alpha P_\alpha$, where
\begin{equation}
(P_\alpha)_{j\beta,l\beta'}=\frac1N\,\delta_{\alpha\beta}\,\delta_{\alpha\beta'}
\cos[q_\alpha(j-l)] ,
\label{eq:mode-projector}
\end{equation}
with weights $\kappa_x=\kappa_y=\kappa_z=\frac12$ if the first entry in \cref{eq:exact-cmc} is the smallest, and $\kappa_\alpha=\kappa_\beta=1$ and zero weight for the remaining component if $s_\alpha+s_\beta$ is the smallest. For $q_\alpha\in\{0,\pi\}$, $P_\alpha$ is the projector onto the Fourier mode $q_\alpha$ of spin component $\alpha$, and for $q_\alpha\notin\{0,\pi\}$ it is one half of the projector onto the two modes $\pm q_\alpha$. Substituting this sum in fact we obtain $\Tr(P_\alpha\Gamma)=\frac4N\Var{J_\alpha(q_\alpha)}$, which afterwards has to be minimized over $q_\alpha$ for the various components.

The proof, which we give in \cref{app:cmc-details}, has three steps. First, averaging an optimal witness over the symmetry transformations of the state, i.e., the lattice translations, the $\pi$ rotation of all spins about the $z$ axis (spin parity), and the reflection $\sigma^y\mapsto-\sigma^y$ of all Bloch vectors (complex conjugation), yields a witness that is equally good and has the same symmetries. Such a witness does not couple different spin components and is diagonal in momentum, so that its value is a weighted sum of the structure factors $\sfac_\alpha(q)$, which can only decrease if every structure factor is replaced by its minimum $s_\alpha$. Second, the separable bound $b(Z)$ of a symmetric witness depends only on the total weights $\kappa_\alpha$ that it assigns to the three spin components, and minimizing $\sum_\alpha\kappa_\alpha s_\alpha$ under the normalization $b(Z)\geq1$ is a linear program in three variables, whose optimal corners give \cref{eq:exact-cmc}. Third, the matrices $P_\alpha$ built from the minimizing Fourier modes attain this bound.

If all three minimizing momenta are $0$ or $\pi$, i.e., they are commensurate, and the triangle inequalities hold, $Z^\star$ is thus one half of a rank-three projector, exactly as found numerically in \cref{fig:cmc-optimal-Z}.  For an incommensurate minimum, i.e., $q_\alpha\notin\{0,\pi\}$, instead, the block $P_\alpha$ has rank two.
At the noncritical point $(\gamma,h)=(0.5,0.5)$, for instance, with $N=10$ and $T=0.1$, the $x$ structure factor is minimal at $q_x=4\pi/5$, the triangle inequality is violated, and the optimum is the two-component witness $s_x+s_z=0.778$.  The numerical optimizer indeed returns a matrix with eigenvalues $(1,\frac12,\frac12)$, in agreement with \cref{thm:reduction}.  More generally, we have checked \cref{eq:exact-cmc} against the numerical SDP at randomly chosen parameters and temperatures, and found agreement to solver precision.

Conversely, the primal side of the SDP gives an explicit decomposition of the covariance matrix into local pure-state covariance matrices mixed with classical correlations (cf.~\cref{cor:certificate} in \cref{app:cmc-details}). The idea is as follows. In a pure state of a single qubit with Bloch vector $\bm r$, the variance of $\sigma^\alpha$ is $1-r_\alpha^2$, so that the local noise in the three spin components adds up to $\sum_\alpha(1-r_\alpha^2)=2$, due to $|\bm r|=1$ holding for all pure states. A separable covariance matrix must contain such noise at every site, cf.\ \cref{eq:cmc-primal}, and all remaining correlations must be classical, i.e., form a positive semidefinite matrix. For the thermal state of the periodic chain we can choose the same local noise $1-x_\alpha$ at every site, with $x_\alpha\geq0$ and $x_x+x_y+x_z=1$, and the remainder is positive semidefinite if and only if
$s_\alpha\geq1-x_\alpha$ holds for every $\alpha$. Such numbers $x_\alpha$ exist if and only if
\begin{equation}
\sum_{\alpha=x,y,z}\max\{0,\,1-s_\alpha\}\leq1 ,
\label{eq:separable-condition}
\end{equation}
which is equivalent to $\omega_{\cmc}(\varrho)\geq1$. The separable covariance matrix is then the sum of the local noise $\bigoplus_j(\id_3-X)$, with $X={\rm diag}(x_x,x_y,x_z)$, and of the positive semidefinite remainder $\bigoplus_\alpha[\Gamma^{\alpha\alpha}-(1-x_\alpha)\id_N]$. See \cref{cor:certificate} in \cref{app:cmc-details} for a more formal proof.

Thus, \cref{thm:reduction} (together with \cref{cor:certificate} in \cref{app:cmc-details} confirming the existence of a separable covariance matrix whenever the witness does not detect entanglement) solves the CMC completely for the thermal XY chain with PBC.
For OBC, the reduction is no longer exact, but the three-mode witnesses remain close to optimal (\cref{app:small-chain-maps}). Moreover, the matrices $Z^\star$ of \cref{thm:reduction} remain admissible witnesses for every state, since $b(Z^\star)=1$ depends only on the weights $\kappa_\alpha$. Evaluated with the structure factors of an open chain, the right-hand side of \cref{eq:exact-cmc} is therefore still the value of a valid CMC witness and thus an upper bound on $\omega_{\cmc}$: it can underestimate, but never overestimate, the violation of the CMC, and values below one certify entanglement.

\subsection{Thermodynamic limit and entanglement temperature}
\label{subsec:thermodynamic-limit}

In this subsection we evaluate the optimal witness in the thermodynamic limit and determine the highest temperature at which it detects entanglement (see also Ref.~\cite{IgloiToth2023} for another investigation of entanglement witnesses of the thermal XY chain). The minimizing momenta at the two reference points, which are the same for $N=10$ and $N\to\infty$ (\cref{tab:structure-factor-minima}), have a simple physical interpretation. At the XX point, the ferromagnetic transverse correlations make $\sfac_x=\sfac_y$ smallest at $q=\pi$, whereas the longitudinal mode at $q=0$ is the conserved magnetization, whose variance is fixed by the uniform susceptibility $\chi=\partial\mean{\sigma^z}/\partial h$ through the fluctuation--dissipation relation,
\begin{equation}
\sfac_z(0)=T\chi ,
\qquad
\omega_{\rm XX}=\sfac_x(\pi)+\tfrac12T\chi ,
\label{eq:xx-susceptibility}
\end{equation}
so that the longitudinal contribution vanishes linearly at low temperature, where $\chi\simeq2/\pi$. This connects the witness to the use of magnetic susceptibilities as entanglement witnesses~\cite{WiesniakVedralBrukner2005}. At the Ising point, the witness avoids the diverging fluctuations $\sfac_x(0)$ of the order parameter and uses instead $\sfac_x(\pi)$ and the uniform $y$ mode, which is conjugate to the order parameter, $\ii[H,\sum_j\sigma_j^x]=2h\sum_j\sigma_j^y$, and has correspondingly suppressed fluctuations, whereas the positive connected longitudinal correlations make the staggered $z$ mode the least fluctuating one.

Because the optimal witness depends on the state only through structure factors, its value can be calculated also in the thermodynamic limit, where the structure factors become
\begin{equation}
\sfac_\alpha(q)=\sum_{r\in\mathbb Z}\e^{\ii qr}
\Bigl[\mean{\sigma_0^\alpha\sigma_r^\alpha}_\beta-\mean{\sigma^\alpha}_\beta^2\Bigr],
\label{eq:structure-factor-tdl}
\end{equation}
which are continuous functions of $q$ and \Cref{thm:reduction} gives the optimal CMC also in the thermodynamic limit.
For the XY chain the correlation functions entering \cref{eq:structure-factor-tdl} are products of fermionic contractions and Toeplitz determinants (\cref{app:xy-solution}), so the optimal CMC of the infinite chain is obtained from a few one-dimensional integrals, without any optimization.  At $T=0.1$, for instance, the minimal structure factors (\cref{fig:structure-factors}) are $s_x=s_y=0.157$ (at $q=\pi$) and $s_z=0.064$ (at $q=0$) at the XX point, and $s_x=0.299$, $s_y=0.453$ and $s_z=0.366$ (at $q=\pi$, $0$ and $\pi$) at the Ising point, so that we find $\omega_{\cmc}=0.189$ at the XX point and $\omega_{\cmc}=0.559$ at the Ising point (\cref{tab:structure-factor-minima}).  At the Ising point the finite chain with $N=10$ is already close to this limit.

\begin{figure*}[t]
\centering
\includegraphics[width=\textwidth]{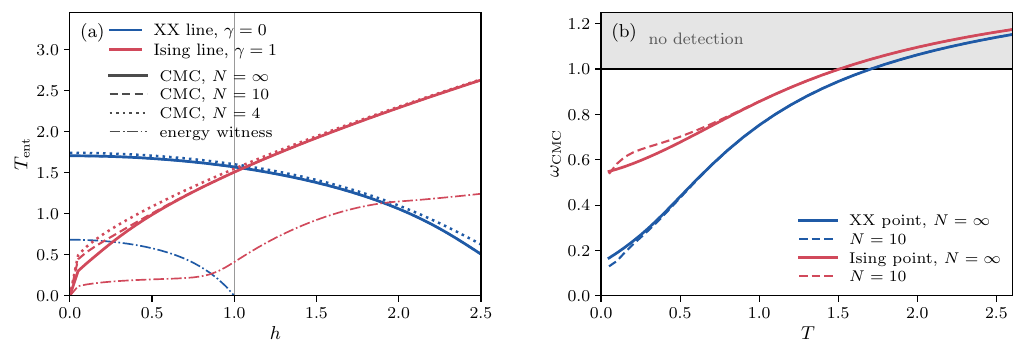}
\caption{(a)~Highest temperature $T_{\rm ent}$ at which the CMC detects entanglement, along the XX line ($\gamma=0$) and the Ising line ($\gamma=1$) for periodic chains with $N=4$ and $N=10$ and in the thermodynamic limit.  Dash-dotted lines: detection threshold of the energy witness in the thermodynamic limit.  The vertical line marks $h=1$, i.e.\ the Ising critical point and the saturation field of the XX chain.  On the Ising line $T_{\rm ent}\to0$ for $h\to0$.  (b)~Optimal CMC value as a function of temperature at the XX and Ising points.  By \cref{prop:bsa-bound} below, $1-\omega_{\cmc}$ is a lower bound on the entangled weight in the best separable approximation.}
\label{fig:entanglement-temperature}
\end{figure*}

The structure-factor formula also makes it possible to calculate the highest temperature at which the CMC detects entanglement, which we denote by $T_{\rm ent}$ and plot in
\cref{fig:entanglement-temperature}
along the XX and Ising lines together with the temperature dependence of $\omega_{\cmc}$ at the two reference points.  In the thermodynamic limit $T_{\rm ent}\simeq1.70$ at the XX point and $T_{\rm ent}\simeq1.51$ at the Ising point.  On the XX line, and on the Ising line for $h\geq1$, for example, the values for $N=10$ agree with the thermodynamic limit to within $2\times10^{-3}$.
For comparison, the energy witness~\cite{DowlingDohertyBartlett2004,Toth2005,IgloiToth2023}, which certifies entanglement whenever the energy lies below the minimum over product states, detects the thermal state only up to $T\simeq0.68$ at the XX point and $T\simeq0.41$ at the Ising point, and not at all on the XX line for $h\geq1$ (\cref{fig:entanglement-temperature}).

Several features are worth noting.  First, the entanglement certified by the covariance matrix survives up to temperatures of the order of the exchange coupling.
Second, $T_{\rm ent}$ is a smooth function of $h$ across the Ising transition at $h=1$: at these temperatures the quantum critical point leaves no sharp trace, and the witness is controlled by the minima of the structure factors, not by their critical peaks.  Third, on the Ising line $T_{\rm ent}$ vanishes for $h\to0$, where the Gibbs state of the classical Ising chain is separable.  In this limit the transverse field reduces the $y$ and $z$ fluctuations by an amount of order $h^2$, while thermally excited domain walls increase the staggered $x$ fluctuations.
Heuristically, the witness is violated as long as
$T\lesssim T_{\rm ent}\simeq1/\ln(1/h)$, up to logarithmic corrections, and numerically we find $T_{\rm ent}\ln(1/h)\simeq0.90$ for $0.05\leq h\leq0.2$.  In the paramagnetic phase, instead, $T_{\rm ent}$ keeps increasing with $h$.
Finally, on the XX line $T_{\rm ent}$ decreases slowly with $|h|$ and remains finite beyond the saturation field $h=1$, where the ground state is fully polarized and thus separable, which is a feature observed also in fully-connected models~\cite{MatheEtAl2026}.  There, $\omega_{\cmc}\to1$ for $T\to0$, and the witness detects the entanglement carried by thermally excited magnons in a window of temperatures below $T_{\rm ent}$, both for finite chains with $N=4$ and $N=10$ and in the thermodynamic limit (\cref{fig:entanglement-temperature}(a)), whereas at very low temperatures no entanglement is detected on this part of the XX line (\cref{fig:witness-map}(a)).

\section{Entanglement generated by a quench from a thermal state}
\label{sec:quench}

Sudden quenches of the XY chain are a typical testbed for non-equilibrium dynamics~\cite{CEF2011,EsslerFagotti2016}: at $t=0$ a parameter is switched instantaneously, $H_i\to H_f$, after which the state evolves unitarily under the time-independent $H_f$. Whenever the initial state is a pure ground state, the final state after unitary evolution will also be a pure state, and the partial transpose can thus detect every state that is entangled across a given cut. It is thus more interesting and realistic to investigate mixed states evolving under unitary evolution, where the partial transpose is no longer a complete criterion. Entanglement witnesses for postquench states of the XY chain, and the comparison between generalized and thermal Gibbs ensembles, have been studied in Refs.~\cite{IgloiToth2023,IgloiKiraly2024}. In this work, we consider quenches starting from thermal states and answer the experimentally motivated question: starting from a thermal state in which no entanglement is detected, can a sudden quench generate entanglement that is certifiable from two-body correlations alone? Such equal-time correlations are routinely measured after quenches in quantum simulators~\cite{Jurcevic2014,Zhang2017}. For bosonic Gaussian systems, the CMC framework has recently shown that unitary compression can generate entanglement from initially separable thermal states~\cite{MatheBose2026}. Here, we positively answer a similar question in the spin-chain analogue, and show how the CMC as a scalable criterion certifies such entanglement.

Because $\varrho(0)$ and $H_f$ are quadratic in Majorana fermions, $\varrho(t)$ remains fermionic Gaussian, and all one- and two-spin correlators follow exactly from Wick's theorem, both for open chains of any length and for the infinite chain (\cref{app:quench}). The quench breaks the symmetry $\varrho=\varrho^*$ of the Gibbs states: since $\varrho(t)^*=\varrho(-t)$, correlations $\langle\sigma^x_i\sigma^y_j\rangle\neq0$ appear for $t>0$. Because of this reduced symmetry, the structure of the optimal CMC witness is more complex, and we simply calculate it numerically from the SDP, which for the infinite chain is still brought to the momentum-block form (cf.~\cref{eq:app-block-sdp} in \cref{app:quench}) by translation invariance.

\begin{figure*}[t]
    \centering
    \includegraphics[width=1.0\linewidth]{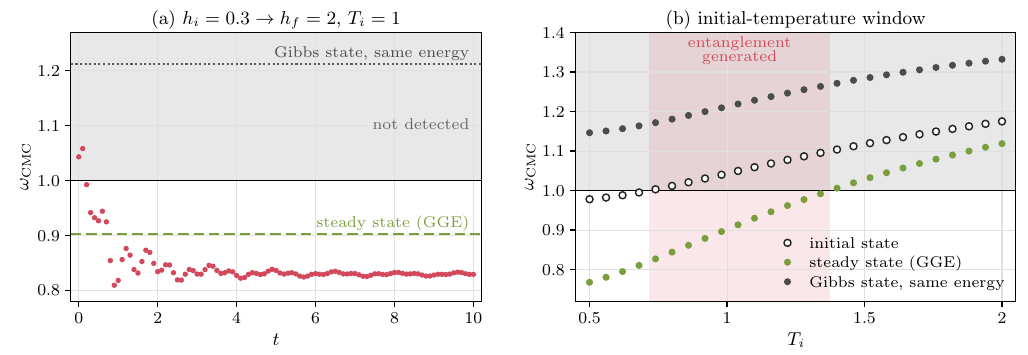}
    \caption{Quench of the infinite Ising chain from $h_i=0.3$ to $h_f=2$. (a)~$\omega_{\cmc}(t)$ for $T_i=1$, with the values of the steady state (GGE) and of the Gibbs state of $H_f$ with the same energy. (b)~$\omega_{\cmc}$ of the initial state, the steady state and the equal-energy Gibbs state versus $T_i$. Shaded red: entanglement generated by the quench. Grey: not detected.}
    \label{fig:quench}
\end{figure*}

We focus on the Ising chain ($\gamma = 1$) and quench the transverse field, i.e., $H_i$ and $H_f$ are given by \cref{eq:xy-hamiltonian} with $\gamma=1$ and with the fields $h_i$ and $h_f$, respectively. The initial state $\varrho(0)$ is the Gibbs state \eqref{eq:thermal-state} of $H_i$ at temperature $T_i$, and $\varrho(t)=\e^{-\ii H_ft}\varrho(0)\,\e^{\ii H_ft}$ for $t>0$. In our study we find that for $T_i=1$, quenches from the ferromagnetic to the paramagnetic regime are the only ones that generate steady-state entanglement from an initially undetected state, apart from a narrow window of weak quenches close to the detection threshold (\cref{fig:quench-app}(a)). \cref{fig:quench}(a) shows the dynamics: The initial state is not detected ($\omega_{\cmc}=1.04$), and from $t\approx0.2$ on, the CMC certifies entanglement at all times. At long times, the state relaxes to a generalized Gibbs ensemble (GGE)~\cite{Rigol2007,CEF2011,EsslerFagotti2016}, for which the SDP gives $\omega_{\cmc}=0.90$. By concavity of $\Gamma$ this value also bounds the time-averaged $\omega_{\cmc}(t)$ from above. By contrast, the Gibbs state of $H_f$ with the same energy ($T_{\rm eff}=4.1$) is far from detection ($\omega_{\cmc}=1.21$). The quench injects energy, and the entanglement survives only because the integrable dynamics retain memory of the initial state. \cref{fig:quench}(b) shows the temperature window. The steady state is certified to be entangled for $T_i<1.37$, and the initial state only for $T_i<0.72$.
The PPT criterion also detects this entanglement, both in blocks of three contiguous spins and in the full state of a chain of $N=8$ spins (\cref{app:quench}). However, no two-body reduced state is entangled at any time. Pairwise tests therefore cannot see this entanglement, and the energy witness fails as well, since the energy $\mean{H_f}$, on which this witness depends, remains constant after the quench and equal to that of the Gibbs state at $T_{\rm eff}$.

In terms of the minimal structure factors \cref{eq:structure-factor-minima}, we find that the ferromagnet suppresses fluctuations of only one spin component ($s_x=0.16$, $s_y=0.98$, $s_z=0.95$). This is compatible with separability. In the steady state, the suppression is shared by all three components ($s_x=0.43$, $s_y=s_z=0.69$ at $q=(\pi,\pi,0)$), which no separable state allows, cf.\ \cref{eq:three-mode-witness}.

\section{Quantifying the detected entanglement}
\label{sec:quantification}

In principle, a witness can provide more than a yes-or-no answer, as its violation can be used to bound an entanglement quantifier.  Following recent results~\cite{FadelEtAl2021,MatheEtAl2026}, in this section we show that the optimal CMC witness bounds the best separable approximation, and, afterwards, also relate this quantity to the geometric measure of entanglement, showing how the same witness can also bound the latter.
Proofs and technical details are given in \cref{app:quantification-details}.

Every state can be decomposed as $\varrho=(1-t)\varrho_{\rm sep}+t\tau$, with $\varrho_{\rm sep}\in\SEP$ and $\tau$ an arbitrary state.  The best separable approximation (BSA) is the decomposition with the largest separable weight~\cite{LewensteinSanpera1998,KarnasLewenstein2001}, and
\begin{equation}
E_{\bsa}(\varrho)=\min\bigl\{t:\ \varrho=(1-t)\varrho_{\rm sep}+t\tau\bigr\}
\label{eq:bsa-definition}
\end{equation}
is the entangled weight that must remain after the largest possible separable part has been removed.  It vanishes exactly on separable states, it equals one on every entangled pure state, and it has the witness representation~\cite{Brandao2005}
\begin{equation}
\begin{split}
E_{\bsa}(\varrho)=\max\bigl\{-\Tr(W\varrho):{}&\ W\succeq-\id,\\
&\ \Tr(W\varrho_{\rm sep})\geq0\ \forall\varrho_{\rm sep}\in\SEP\bigr\}.
\end{split}
\label{eq:bsa-witness-dual}
\end{equation}
Every witness with $W\succeq-\id$ therefore yields a lower bound.  The CMC witnesses are nonlinear in $\varrho$, but they can be linearized around the measured first moments.

\begin{proposition}[BSA from the covariance witness]
\label{prop:bsa-bound}
For every $Z\succeq0$ with $b(Z)=1$,
\begin{equation}
E_{\bsa}(\varrho)\geq\max\bigl\{0,\,1-\Tr[Z\Gamma(\varrho)]\bigr\}.
\label{eq:bsa-bound}
\end{equation}
In particular $E_{\bsa}(\varrho)\geq 1-\omega_{\cmc}(\varrho)$, and from \cref{thm:reduction} we find a bound as an explicit function of the three minimal structure factors for the XY model at equilibrium.
\end{proposition}

\begin{proof}
We follow the arguments of Refs.~\cite{FadelEtAl2021,MatheEtAl2026}. Write $Z=\sum_\nu z_\nu\bm v_\nu\bm v_\nu^\T$ with the observables $A_\nu$ of \cref{eq:cmc-lur}. Since every variance is a minimum of linear expectation values, $\Var{A}_\varrho=\min_{a\in\mathbb R}\mean{(A-a\id)^2}_\varrho$, attained at $a=\mean{A}_\varrho$, we can write
\begin{equation}
\begin{gathered}
\Tr[Z\Gamma(\varrho)]-1=\min_{\bm a}\Tr(W_{\bm a}\varrho),
\\
W_{\bm a}=\sum_\nu z_\nu(A_\nu-a_\nu\id)^2-\id ,
\end{gathered}
\label{eq:linearized-witness}
\end{equation}
with $\bm a=(a_\nu)$. For every fixed $\bm a$, $W_{\bm a}$ is a linear entanglement witness, since $\Tr(W_{\bm a}\varrho_{\rm sep})$ is bounded from below by $\Tr[Z\Gamma(\varrho_{\rm sep})]-1$, which is nonnegative for all $\varrho_{\rm sep}\in\SEP$ by \cref{eq:cmc-dual-bound}, and it is normalized such that $W_{\bm a}\succeq-\id$. Hence, for every decomposition $\varrho=(1-t)\varrho_{\rm sep}+t\tau$ as in \cref{eq:bsa-definition},
\begin{equation*}
\Tr(W_{\bm a}\varrho)=(1-t)\Tr(W_{\bm a}\varrho_{\rm sep})+t\Tr(W_{\bm a}\tau)\geq-t .
\end{equation*}
Choosing $a_\nu=\mean{A_\nu}_\varrho$, which attains the minimum in \cref{eq:linearized-witness}, we obtain $t\geq1-\Tr[Z\Gamma(\varrho)]$ for every such decomposition, and hence \cref{eq:bsa-bound}, since $E_{\bsa}\geq0$.
\end{proof}

The BSA can be understood as a geometric measure, i.e., as some distance with respect to the set of fully separable states, similarly to the geometric measure of entanglement $E_G(\varrho)$~\cite{WeiGoldbart2003}. Moreover, both $E_{\bsa}$ and $E_G(\varrho)$ are convex roofs,
\begin{equation}
\begin{aligned}
    E_{\bsa}(\varrho)&=\inf_{\{p_k,\psi_k\}}\sum_kp_k\,e(\psi_k),
\\
E_G(\varrho)&=\inf_{\{p_k,\psi_k\}}\sum_kp_k\bigl[1-\Lambda^2(\psi_k)\bigr],
\label{eq:convex-roofs}
\end{aligned}
\end{equation}
where the infima run over all pure-state decompositions $\varrho=\sum_kp_k\ketbra{\psi_k}$, $e(\psi)$ equals zero for product states and one otherwise, and $\Lambda^2(\psi)=\max_{\phi}|\braket{\phi}{\psi}|^2$ is the largest overlap with a product state (\cref{app:bsa-geometric-roofs}).  The BSA thus counts the weight of the entangled components of the optimal ensemble, while the geometric measure weights each of them by its distance from the set of product states.  In this sense $E_G$ is a geometrically weighted BSA, and
\begin{equation}
0\leq E_G(\varrho)\leq E_{\bsa}(\varrho)\leq1 .
\label{eq:geometric-bsa-order}
\end{equation}
Moreover, $E_G(\varrho)=1-\max_{\varrho_{\rm sep}\in\SEP}F(\varrho,\varrho_{\rm sep})$, with $F$ the squared Uhlmann fidelity~\cite{StreltsovKampermannBruss2010}, so that any explicit separable state provides an upper bound.  The geometric measure has been studied extensively for ground states of spin chains, where it signals quantum criticality~\cite{WeiEtAl2005,Orus2008}.

Because of \cref{eq:geometric-bsa-order}, \cref{prop:bsa-bound} does not bound $E_G$ from below.  A lower bound on a convex-roof measure from a measured witness value $w=\Tr(W\varrho)$ can be instead obtained from its Legendre transform~\cite{GuhneReimpellWerner2007,GuhneReimpellWerner2008},
\begin{equation}
E_G(\varrho)\geq\sup_{\lambda\in\mathbb R}\Bigl\{1+\lambda w-
\max_{\ket\phi\ {\rm product}}\lambda_{\max}\bigl(\lambda W+\ketbra\phi\bigr)\Bigr\},
\label{eq:geometric-legendre-bound}
\end{equation}
where we have evaluated the conjugate function of the geometric measure explicitly (\cref{app:geometric-legendre-details}).  This is a (hard) problem that can be treated numerically. However, the difficult part is the inner maximization over product states, which for our witness simplifies considerably.  In particular, in the cases treated here the linearized witnesses of \cref{eq:linearized-witness} can be brought via local unitaries to
$W=\frac2N\bm J^2-\id$, i.e., to a function of the total spin only. What is left is thus maximizing the largest eigenvalue of $\lambda W+\ketbra\phi$ over all product states $\ket\phi$, which can be potentially also relaxed by considering states with a positive partial transpose across every bipartition.  Since the witness is invariant under permutations and collective rotations, this relaxation reduces to a small linear program over the weights of the total-spin sectors, so that a lower bound on $E_G$ can be calculated from the measured witness value alone. We describe this calculation in \cref{app:geometric-legendre-details}, together with a numerical search over product states, which suggests that the inner maximum is attained at Néel-type states.

Complementary upper bounds on $E_{\bsa}$ and $E_G$ can be obtained from explicit separable states, e.g.\ from product-state ensembles symmetrized with the symmetry group of the thermal state and further optimized~\cite{KarnasLewenstein2001,OhstEtAl2024,MatheEtAl2026}, or from flat solutions of moment relaxations for the separable part of the decomposition in \cref{eq:bsa-definition}, analogous to the hierarchy of \cref{subsec:entanglement-framework}. Finding such decompositions is, however, a hard problem: deciding whether a given state is separable is NP-hard in general~\cite{Gurvits2003,Gharibian2010}, and even for permutationally invariant states, whose symmetry drastically reduces the number of free parameters, explicit separable decompositions remain demanding to construct~\cite{Tura2018,MatheEtAl2026}.  We leave a systematic study of these explicit separable decompositions for future work.

Finally, the CMC has also been generalized to bound the Schmidt number, i.e., the entanglement dimensionality across a bipartition~\cite{LiuEtAl2024}. Such bounds are not directly accessible from the witnesses derived here, which test full separability with single-spin observables, but they could be obtained by applying these generalized criteria to covariance matrices of observables acting on blocks of spins on both sides of a bipartition, which we leave for future work as well.

\section{Conclusions and outlook}
\label{sec:conclusions}

We have used the thermal XY chain to ask what the optimal covariance-matrix entanglement witness looks like, and what it reveals about the equilibrium physics of the model.  The small-chain benchmark shows that the covariance-matrix criterion, which uses only one- and two-spin correlations, is nearly as powerful as a full moment-matrix hierarchy around the XX and Ising critical regions, and that it detects entanglement in states without pairwise entanglement and, close to the XX point, even in states that are PPT across every bipartition.

The central result is an exact solution of the CMC for the thermal states of the periodic XY chain.  Owing to their invariance under translations, spin parity and complex conjugation, the optimal covariance witness is fixed by the minima over momentum of the three static spin structure factors, and whenever this witness is satisfied the covariance matrix of the thermal state admits an explicit decomposition of separable form.  The optimal witness selects the least fluctuating collective modes, so that its form is dictated by symmetry rather than by criticality.  In the thermodynamic limit the witness yields entanglement temperatures of the order of the exchange coupling, well above those of the energy witness.  The same witness bounds the entangled weight in the best separable approximation and the geometric measure of entanglement.  Out of equilibrium, the CMC certifies the entanglement generated by a quench from a thermal state in which no entanglement is detected, even though no pair of spins becomes entangled.

Several extensions are natural, e.g., to other spin models, such as XXZ and XYZ chains, for which structure factors are known from Bethe ansatz, numerics or neutron scattering, and to lattices in higher dimensions.  It would also be interesting to study the scaling of the minimal structure factors, and hence of the witness, in the quantum-critical regime. Moreover, the covariance witnesses could be extended to stronger forms of entanglement: to genuine multipartite entanglement, which in spin chains has been certified, e.g., with energy witnesses and PPT mixtures~\cite{GuhneTothBriegel2005,JungnitschMoroderGuhne2011,HofmannOsterlohGuhne2014} and which would require a bound for biseparable states (\cref{app:gme}), and to high-dimensional entanglement across bipartitions of the chain, as quantified by the Schmidt number, for which covariance-matrix criteria are available~\cite{LiuEtAl2024,Liu_2026_multipartite}. Finally, higher levels of the moment hierarchy with flat extensions could determine where the thermal state becomes provably separable.

\section*{Acknowledgements}

The authors used ChatGPT~5.6 and Claude Opus~5.5 for assistance with coding, graphics, symbolic calculations, and the editing of the text. The authors provided the main ideas and proof directions and checked all results, taking full responsibility for the content of the article.

This research was funded in whole or in part by the Austrian Science Fund (FWF) [\href{https://doi.org/10.55776/P35810}{10.55776/P35810}], and [\href{https://doi.org/10.55776/P36633}{10.55776/P36633}]. For open access purposes, the authors have applied a CC BY public copyright license to any author-accepted manuscript version arising from this submission. G.V. also acknowledges support from the Grant No. RYC2024-048278-I funded by MCIU/AEI/10.13039/501100011033 and FSE+. A.U. is financially supported by JSPS Overseas Research Fellowships and acknowledges support from grant PID2022-139099NB-I00 funded by MCIN/AEI/10.13039/501100011033 with the support of FEDER funds. O.G. is supported by the Deutsche Forschungsgemeinschaft (DFG, German Research Foundation, project number 563437167) and the German Federal Ministry of Research, Technology and Space (Projects BeRyQC, Grant No.\ 01N124023D, and PraQTisch, Grant No.\ 01N126223D).

\FloatBarrier
\appendix

\section{Free-fermion solution and thermal correlations}
\label{app:xy-solution}

This appendix fixes our conventions and summarizes the calculation of the thermal spin correlators. Full derivations can be found in Refs.~\cite{JordanWigner1928,LiebSchultzMattis1961,Pfeuty1970, BarouchMcCoyDresden1970,BarouchMcCoy1971}.

The Jordan--Wigner transformation introduces spinless fermions
\begin{equation}
c_j=\Bigl(\prod_{l<j}\sigma_l^z\Bigr)\sigma_j^- ,
\qquad
c_j^\dagger=\Bigl(\prod_{l<j}\sigma_l^z\Bigr)\sigma_j^+ ,
\label{eq:jw}
\end{equation}
with $\sigma_j^z=2c_j^\dagger c_j-1$.  For OBC the Hamiltonian \cref{eq:xy-hamiltonian} becomes
\begin{equation}
\begin{split}
H={}&\sum_{j=1}^{N-1}\Bigl[c_j^\dagger c_{j+1}+c_{j+1}^\dagger c_j
+\gamma\bigl(c_j^\dagger c_{j+1}^\dagger-c_jc_{j+1}\bigr)\Bigr]\\
&-2h\sum_{j=1}^N\Bigl(c_j^\dagger c_j-\frac12\Bigr) .
\end{split}
\label{eq:fermionic-xy-obc}
\end{equation}
For PBC the bond between sites $N$ and $1$ contains the spin parity, and the Hamiltonian is diagonalized separately in the sectors $p=\pm1$ of $P_z$.  For even $N$, the sector $p=+1$ uses antiperiodic momenta $K_+=\{(2m+1)\pi/N\}$ and the sector $p=-1$ periodic momenta $K_-=\{2m\pi/N\}$, with the unpaired modes $k=0,\pi$ treated separately.

In each sector, a Fourier transform, $c_j=N^{-1/2}\sum_{k\in K_p}\e^{-\ii kj}c_k$, followed by a Bogoliubov transformation that mixes $c_k$ and $c_{-k}^\dagger$, brings the Hamiltonian into the diagonal form $H=\sum_k\varepsilon_k(d_k^\dagger d_k-\frac12)$ of \cref{subsec:xy-model}, with the dispersion \cref{eq:dispersion} and the Bogoliubov angle $\theta_k$ given by
\begin{equation}
\cos\theta_k=\frac{2}{\varepsilon_k}\Bigl(\cos k-h\Bigr),
\qquad
\sin\theta_k=\frac{2}{\varepsilon_k}\,\gamma\sin k\,.
\label{eq:bogoliubov-angle}
\end{equation}
Since the quasiparticles have Fermi--Dirac occupations $\mean{d_k^\dagger d_k}_\beta=(\e^{\beta\varepsilon_k}+1)^{-1}$ in the thermal state, the fermionic correlations are
\begin{equation}
\begin{aligned}
\mean{c_k^\dagger c_k}_\beta&=\frac12\Bigl[1-\cos\theta_k\tanh\frac{\beta\varepsilon_k}{2}\Bigr],\\
\mean{c_kc_{-k}}_\beta&=-\frac\ii2\sin\theta_k\tanh\frac{\beta\varepsilon_k}{2},
\end{aligned}
\label{eq:fermion-correlations}
\end{equation}
which hold also for the unpaired modes $k=0,\pi$, with $\sin\theta_k=0$. For a finite periodic chain, physical thermal averages are obtained by combining the two sectors with the parity projectors $(\id\pm P_z)/2$.
For OBC, one instead diagonalizes the $2N\times2N$ Bogoliubov--de Gennes matrix.

The spin-spin correlators can be then computed from the fermionic correlators via Wick's theorem, which results in
\begin{align}
\mean{\sigma_i^z}_\beta&=-G_{ii},
\label{eq:z-G}\\
\mean{\sigma_i^z\sigma_j^z}_\beta&=G_{ii}G_{jj}-G_{ij}G_{ji},
\qquad i\neq j,
\label{eq:zz-G}\\
\mean{\sigma_i^x\sigma_{i+r}^x}_\beta&=\det\bigl[G_{i+a-1,i+b}\bigr]_{a,b=1}^{r},
\label{eq:xx-G}\\
\mean{\sigma_i^y\sigma_{i+r}^y}_\beta&=\det\bigl[G_{i+a,i+b-1}\bigr]_{a,b=1}^{r} ,
\label{eq:yy-G}
\end{align}
where $G_{mn}=-\mean{B_mA_n}_\beta$ is the correlation matrix of the Majorana operators $A_j=c_j^\dagger+c_j$ and $B_j=c_j^\dagger-c_j$. By Fourier transforming \cref{eq:fermion-correlations}, one finds for the infinite chain, which is translation invariant, $G_{mn}=G_{n-m}$ with
\begin{equation}
\begin{split}
G_r=\int_0^\pi\frac{\dd k}{\pi}\,
\frac{2}{\varepsilon_k}&\Bigl[\Bigl(\cos k-h\Bigr)\cos(kr)
+\gamma\sin k\sin(kr)\Bigr]\\
&\times\tanh\frac{\beta\varepsilon_k}{2}.
\end{split}
\label{eq:G-integral}
\end{equation}
For finite periodic chains, the integral is replaced by the corresponding sum over the momenta $K_p$ of each parity sector, and for OBC $G$ follows from the eigenvectors of the Bogoliubov--de Gennes matrix.

Mixed transverse--longitudinal correlators vanish by spin parity, and the real equilibrium state has no symmetrized $xy$ correlator.  These expressions thus provide all entries of the covariance matrix.
In particular, the magnetization of the infinite chain is
\begin{equation}
\mean{\sigma^z}_\beta=-G_0=-\int_0^\pi\frac{\dd k}{\pi}\,\cos\theta_k\tanh\frac{\beta\varepsilon_k}{2},
\label{eq:magnetization-integral}
\end{equation}
and \cref{eq:zz-G} gives the longitudinal structure factor in closed form,
\begin{equation}
\sfac_z(q)=1-G_0^2-\sum_{r\neq0}\e^{\ii qr}\,G_rG_{-r} .
\label{eq:Sz-closed-form}
\end{equation}
On the XX line, $\cos\theta_k=\operatorname{sgn}(\cos k-h)$, so that $\mean{\sigma^z}_\beta=\int_0^\pi\frac{\dd k}{\pi}\tanh[\beta(h-\cos k)]$, and the susceptibility in \cref{eq:xx-susceptibility} is
\begin{equation}
\begin{split}
\chi=\frac{\partial\mean{\sigma^z}_\beta}{\partial h}&=\int_0^\pi\frac{\dd k}{\pi}\,\frac{\beta}{\cosh^2[\beta(h-\cos k)]}\\
&\xrightarrow{\;T\to0\;}\;\frac{2}{\pi\sqrt{1-h^2}}\qquad(|h|<1),
\end{split}
\label{eq:xx-chi-integral}
\end{equation}
i.e., $\chi\simeq2/\pi$ at $h=0$. Since the magnetization is conserved for $\gamma=0$, $\sfac_z(0)$ from \cref{eq:Sz-closed-form} equals $T\chi$.

In the thermodynamic-limit
results of \cref{subsec:thermodynamic-limit} we evaluate \cref{eq:G-integral} by quadrature and sum \cref{eq:structure-factor-tdl} up to distances at which the correlations have decayed below $10^{-9}$, or up to $400$ sites.  Deep in the ordered phase the thermal correlation length can exceed this range.  There we evaluate the structure factors only at the commensurate momenta $0$ and $\pi$, which are the minima on the Ising line, and we accelerate the convergence of the alternating series for the staggered transverse structure factors by repeated averaging of partial sums.  We proceed in the same way for the low-temperature maps of \cref{fig:witness-map}.

The energy witness used for comparison in \cref{subsec:thermodynamic-limit} follows from the same contractions. The energy per site of the infinite chain is
\begin{equation}
e(T)=-\Bigl[\frac{1+\gamma}{2}\,G_1+\frac{1-\gamma}{2}\,G_{-1}\Bigr]+hG_0 ,
\label{eq:energy-per-site}
\end{equation}
and for $\gamma\geq0$ its minimum over product states is $e_{\rm sep}=-(1+\gamma)/2-h^2/[2(1+\gamma)]$ for $|h|\leq 1+\gamma$ and $e_{\rm sep}=-|h|$ otherwise. The energy witness detects entanglement whenever $e(T)<e_{\rm sep}$~\cite{DowlingDohertyBartlett2004,Toth2005}.

\section{Equilibrium regimes of the XY chain}
\label{app:critical}

Here we give more precise statements on the equilibrium regimes of the XY chain summarized in \cref{subsec:xy-model}. Derivations and further details can be found, e.g., in Refs.~\cite{LiebSchultzMattis1961,Pfeuty1970,BarouchMcCoy1971,Sachdev2011,DuttaEtAl2015,Franchini2017}. We consider $\gamma\geq0$ and $h\geq0$, since the other cases follow by symmetry.

The regimes can be read off from the single-fermion dispersion $\varepsilon_k$ of \cref{eq:dispersion}, whose minimum $\Delta=\min_k\varepsilon_k$ is the energy gap of the infinite chain. For $\gamma>0$ and $h$ close to one, the minimum lies at $k=0$, and $\Delta=2|1-h|$ closes linearly at the Ising transition $h=1$. Around this point, $\varepsilon_k^2\simeq\Delta^2+v^2k^2$ for small $k$, with the velocity $v=2\gamma$, so that the low-energy excitations have a relativistic dispersion, with dynamical critical exponent $z=1$. Correspondingly, the correlation length $\xi$ of the ground state, defined below, diverges as $\xi\propto|1-h|^{-\nu}$ with $\nu=1$. For $h<1$ the order parameter is the spontaneous magnetization $m_x$, whose square is the long-distance limit of the ground-state correlations of the infinite chain~\cite{BarouchMcCoy1971},
\begin{equation}
m_x^2=\lim_{r\to\infty}\mean{\sigma_0^x\sigma_r^x}
=\frac{2\sqrt\gamma}{1+\gamma}\Bigl[1-h^2\Bigr]^{1/4},
\label{eq:app-magnetization}
\end{equation}
so that $m_x\propto(1-h)^{1/8}$ vanishes at the transition with the order-parameter exponent $1/8$~\cite{BelavinPolyakovZamolodchikov1984,Cardy1996,DiFrancesco1997,Mussardo2010}.
The low-energy physics is described by the conformal field theory with central charge $c=1/2$ of a free Majorana fermion~\cite{BelavinPolyakovZamolodchikov1984,Cardy1996,DiFrancesco1997,Mussardo2010}.

For $\gamma=0$, the pairing terms of the fermionic Hamiltonian \cref{eq:fermionic-xy-obc}, which are proportional to $\gamma$, vanish, and the magnetization is conserved. The model then describes free fermions with the single-particle energies $2\cos k-2h$ in a band that is partially filled for $h<1$, with Fermi momenta $\pm k_{\rm F}$, $\cos k_{\rm F}=h$. The spectrum $\varepsilon_k=2|\cos k-h|$ is gapless, with linear dispersion and velocity $v=2\sqrt{1-h^2}$ around the Fermi points, so that again $z=1$. The ground-state correlations decay algebraically with distance: the transverse ones as $\mean{\sigma_0^x\sigma_r^x}\propto r^{-1/2}$~\cite{McCoy1968,BarouchMcCoy1971}, i.e., with $x_\sigma=1/4$, and the connected longitudinal ones as $\mean{\sigma_0^z\sigma_r^z}-\mean{\sigma^z}^2=-4\sin^2(k_{\rm F}r)/(\pi r)^2$, as follows from \cref{eq:zz-G}.
The low-energy physics is that of a Luttinger liquid, i.e., of a conformal field theory with $c=1$~\cite{Haldane1981,Giamarchi2003}. A small anisotropy opens a gap that grows linearly with $\gamma$, to leading order $\Delta\simeq2\gamma\sqrt{1-h^2}$, e.g., $\Delta=2\gamma$ at $h=0$ for all $\gamma\leq1$. At the endpoint $(\gamma,h)=(0,1)$ of the critical segment, the two Fermi points merge at $k=0$, the dispersion becomes quadratic, $\varepsilon_k\simeq k^2$, and $z=2$. For $h\geq1$ the ground state of the XX chain is fully polarized and hence a product state.

Inside the ordered region, on the circle $h^2+\gamma^2=1$, the ground state is a product state~\cite{KurmannThomasMuller1982,MullerShrock1985,GiampaoloAdessoIlluminati2008}. This circle separates the region outside it, in which the ground-state correlations approach their limit at large distances monotonically, from the region inside it, in which they oscillate with a wave vector that changes continuously with $\gamma$ and $h$ and is in general incommensurate with the lattice~\cite{BarouchMcCoy1971}.

We define the correlation length $\xi$ through the exponential approach of the correlations to their value at infinite distance,
\begin{equation}
\frac1\xi=-\limsup_{r\to\infty}\frac1r\ln\Bigl|\mean{\sigma_0^\alpha\sigma_r^\alpha}
-\lim_{r'\to\infty}\mean{\sigma_0^\alpha\sigma_{r'}^\alpha}\Bigr| ,
\label{eq:app-correlation-length}
\end{equation}
i.e., the difference decays as $\e^{-r/\xi}$, up to algebraic prefactors and, inside the circle, oscillating factors. In general, $\xi$ depends on the spin component $\alpha$. At $T=0$ it diverges at the Ising transitions, and on the critical XX segment the correlations decay algebraically, i.e., $\xi=\infty$.

At $T>0$, the one-dimensional chain with short-range interactions has no phase transition. Thermal fluctuations destroy long-range order, $\lim_{r\to\infty}\mean{\sigma_0^\alpha\sigma_r^\alpha}=\mean{\sigma^\alpha}^2$, and $\xi$ is finite for all $\gamma$ and $h$. Nevertheless, the ground-state phases leave clear traces at low temperatures~\cite{Sachdev2011}. Where the spectrum has a gap $\Delta$, the excitations are thermally activated, with a density suppressed as $\e^{-\Delta/T}$, and the correlations remain close to those of the ground state, up to this dilute gas of thermal quasiparticles. In the ordered phase, the quasiparticles are domain walls of the order parameter, and their typical distance sets the correlation length of the order-parameter correlations, which therefore grows exponentially as $T\to0$. For the Ising chain, for instance, $\xi^{-1}\simeq\int_{-\pi}^{\pi}\frac{\dd k}{\pi}\,\e^{-\varepsilon_k/T}$ at low temperatures~\cite{SachdevYoung1997,Sachdev2011}. Above a quantum critical point, instead, the temperature is the only energy scale, and it sets the correlation length itself, $\xi\propto T^{-1/z}$. At the critical points with $z=1$, conformal invariance gives $\mean{\sigma_0^x\sigma_r^x}\propto[(\pi T/v)/\sinh(\pi Tr/v)]^{2x_\sigma}$ at low temperatures and large distances, i.e., $\xi=v/(2\pi x_\sigma T)$, which yields $\xi=4v/(\pi T)$ at the Ising transitions and $\xi=2v/(\pi T)$ on the critical XX segment, in agreement with the exact results~\cite{ItsEtAl1993,SachdevYoung1997}. This quantum-critical regime fans out from the critical points as the temperature increases, up to temperatures of the order of the exchange coupling, at which the lattice dispersion becomes important~\cite{Sachdev2011}.

\section[\texorpdfstring{Details on the moment-matrix hierarchy}{Details on the moment-matrix hierarchy}]{Details on the moment-matrix hierarchy}
\label{app:moment-matrix}

In this appendix we give the explicit form of the moment-matrix hierarchy of \cref{subsec:entanglement-framework} and show that its first level is equivalent to the CMC if only one- and two-spin correlations are supplied.

It can be shown~\cite{BohnetWaldraffBraunGiraud2017} that the relaxation of \cref{eq:pauli-classical-moments} needs the positivity conditions on the moment matrix and on the so-called localizing matrices
\begin{equation}
M_k(y)\succeq0,
\qquad
M_{k-1}\!\left[(1-|\bm r_j|^2)\,y\right]\succeq0
\quad(j=1,\dots,N),
\label{eq:moment-matrix}
\end{equation}
at relaxation order $k$ (see also \cite{Lasserre2001,Parrilo2003,Laurent2009,Lasserre2010}). Here, we call $y=(y_{\bm a})$ the sequence of moments $y_{\bm a}=\int\bm x^{\bm a}\,\dd\mu(\bm x)$ of the monomials $\bm x^{\bm a}=\prod_{j=1}^N\prod_{\alpha=x,y,z}(r_j^\alpha)^{a_{j\alpha}}$ of degree $|\bm a|=\sum_{j,\alpha}a_{j\alpha}\leq2k$, with $y_{\bm0}=1$. The moment matrix $M_k(y)$, whose entries are the moments, is correspondingly indexed by the monomials of degree at most $k$, $[M_k(y)]_{\bm a\bm b}=y_{\bm a+\bm b}$, and the localizing matrices $M_{k-1}\!\left[(1-|\bm r_j|^2)\,y\right]$ by those of degree at most $k-1$, $[M_{k-1}((1-|\bm r_j|^2)y)]_{\bm a\bm b}=y_{\bm a+\bm b}-\sum_\alpha y_{\bm a+\bm b+2\bm e_{j\alpha}}$, where $\bm e_{j\alpha}$ is the multi-index of the monomial $r_j^\alpha$. Both matrices are linear in $y$, and \cref{eq:moment-matrix} is necessary for $y$ to be the moment sequence of a measure on the Bloch balls: for every polynomial $p(\bm x)=\sum_{\bm a}p_{\bm a}\bm x^{\bm a}$ of degree at most $k$ one has $\bm p^\T M_k(y)\,\bm p=\int p^2\,\dd\mu\geq0$, and similarly $\int(1-|\bm r_j|^2)\,p^2\,\dd\mu\geq0$ for degree at most $k-1$. Through \cref{eq:pauli-classical-moments}, the data fix the moments of the monomials that are products of components of distinct spins, whereas all other moments, e.g., those of $(r_j^x)^2$ or $r_j^xr_j^y$, are free variables of the relaxation. Each order $k$ thus defines a semidefinite feasibility problem, and the conditions become more restrictive as $k$ increases.

If only one- and two-spin correlations are supplied, the first level of the hierarchy is equivalent to the CMC. At order $k=1$ the monomials are $1$ and the $3N$ components $r_j^\alpha$, so that, generalizing \cref{eq:moment-matrix-example},
\begin{equation}
M_1(y)=\begin{pmatrix}1&\bm m^\T\\ \bm m&C\end{pmatrix},
\label{eq:first-moment-matrix}
\end{equation}
whose entries are the moments $y$, namely the measured means $m_{j\alpha}=\mean{\sigma_j^\alpha}_\varrho$ and the second moments $C_{j\alpha,l\beta}=\int r_j^\alpha r_l^\beta\,\dd\mu$. For $j\neq l$, the latter are the measured correlations $\mean{\sigma_j^\alpha\sigma_l^\beta}_\varrho$, whereas the local second moments $(X_j)_{\alpha\beta}=C_{j\alpha,j\beta}$ are not fixed by the data, since \cref{eq:pauli-classical-moments} holds only for distinct spins. The localizing conditions reduce to $\Tr X_j\leq1$, and $M_1(y)\succeq0$ is equivalent to $C-\bm m\bm m^\T\succeq0$. The off-diagonal blocks of $C-\bm m\bm m^\T$ are those of $\Gamma(\varrho)$, and its diagonal blocks are $X_j-\bm m_j\bm m_j^\T=\Gamma_{jj}-(\id_3-X_j)$, since $\Gamma_{jj}=\id_3-\bm m_j\bm m_j^\T$, where $\bm m_j$ collects the three means $m_{j\alpha}$ of spin $j$. In other words, the Schur complement of $M_1$ with respect to its first entry is $\Gamma(\varrho)-\bigoplus_j(\id_3-X_j)$, where $X_j$ are the unconstrained local second moments, whose trace is at most one by the localizing constraint and can be set to one without loss of generality, because adding a multiple of $\id_3$ to $X_j$ preserves the positivity of the Schur complement. The first level is thus feasible if and only if $\Gamma(\varrho)$ is a separable covariance matrix in the sense of \cref{eq:cmc-primal}.  Higher levels and higher-order correlations make the hierarchy stronger.

\section[Complementary numerical results]{Complementary numerical results}
\label{app:small-chain-maps}

\begin{figure*}[t]
\centering
\includegraphics[width=\textwidth]{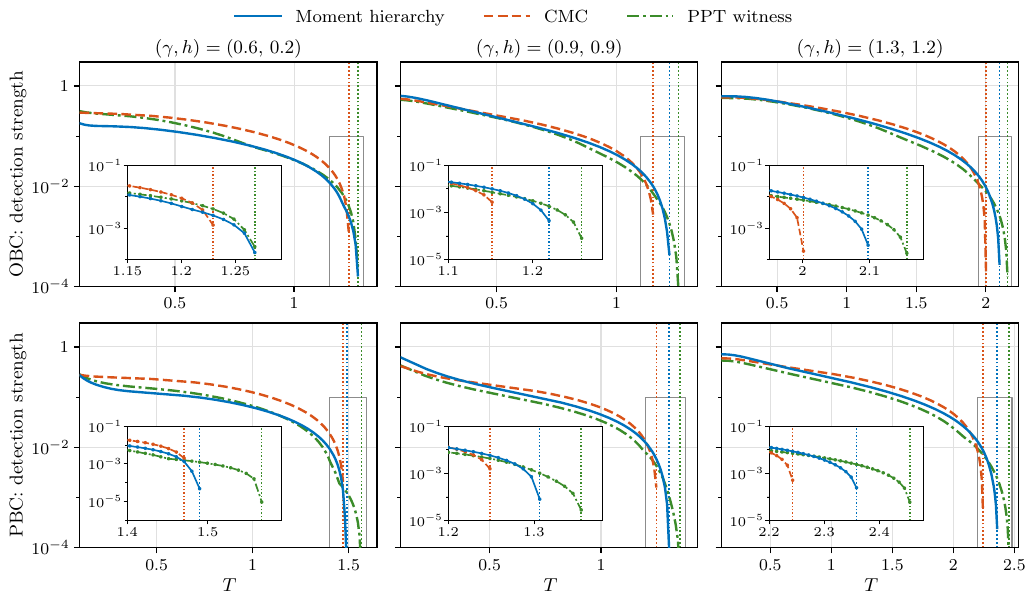}
\caption{Same as \cref{fig:criteria-comparison} at three further points of the $(\gamma,h)$ plane: $(0.6,0.2)$ inside the factorizing circle, $(0.9,0.9)$ in the ordered phase close to the Ising transition, and $(1.3,1.2)$ in the paramagnetic phase. At all three points the PPT witness detects entanglement up to temperatures at least as high as the moment hierarchy, which in turn exceeds the CMC.}
\label{fig:criteria-comparison-app}
\end{figure*}

\begin{figure*}[t]
\centering
\begin{subfigure}[b]{1.0\textwidth}
\includegraphics[width=\linewidth]{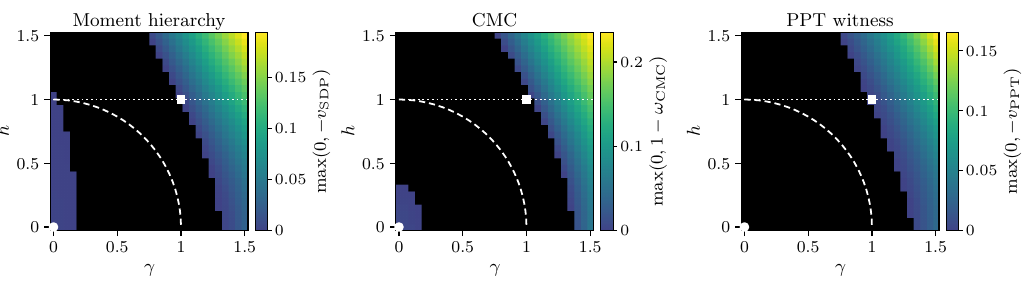}
\caption{Open boundary conditions.}
\end{subfigure}\vfill
\begin{subfigure}[b]{1.0\textwidth}
\includegraphics[width=\linewidth]{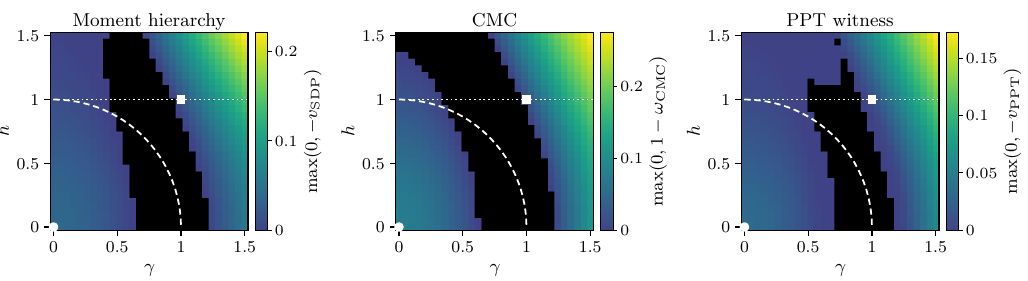}
\caption{Periodic boundary conditions.}
\end{subfigure}
\caption{Entanglement detection in the $(\gamma,h)$ plane at $T\simeq1.47$ for a chain of $N = 4$ sites with open and periodic boundary conditions: Moment hierarchy (left), CMC witness (center), and PPT witness for the alternating cut (right). On this grid, the alternating cut is the most negative of all seven bipartitions wherever any partial transpose is negative. Black: not detected.}
\label{fig:1p47}
\end{figure*}

\begin{figure*}[t]
    \centering
    \includegraphics[width=0.8\linewidth]{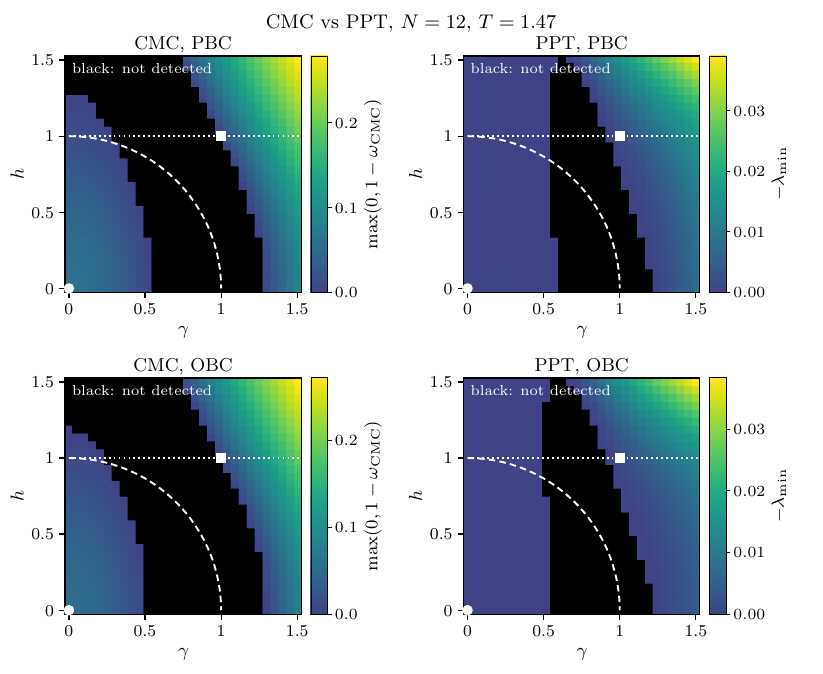}
    \caption{CMC (left) and PPT test for the alternating cut (right) in the $(\gamma,h)$ plane at $T=1.47$ for $N=12$, with periodic (top) and open (bottom) boundary conditions. The colors show $\max(0,1-\omega_{\cmc})$ and the modulus of the smallest eigenvalue of the partial transpose, respectively, and black points are not detected. Dashed line: factorizing circle; dotted line: Ising critical field; circle and square: XX and Ising points.}
    \label{fig:N_12}
\end{figure*}

Here we collect numerical results that complement \cref{sec:small-chain-results,sec:analytic-witness}. \Cref{fig:criteria-comparison-app} extends \cref{fig:criteria-comparison} to three further points of the $(\gamma,h)$ plane. \Cref{fig:1p47} compares the three tests of \cref{sec:small-chain-results} in the $(\gamma,h)$ plane at $T=1.47$ for $N=4$.
For OBC, close to the boundary of the detection region at large $\gamma$, the PPT test detects a few points missed by the hierarchy. For PBC, the PPT test detects every point detected by the moment hierarchy and the CMC, and many more, but at the XX point the window of fully PPT states detected by the CMC opens at slightly higher temperatures, $1.728<T<1.740$ (\cref{fig:criteria-comparison}). This window persists in longer chains. Checking all bipartitions, we find it for $1.616<T<1.659$ (OBC) and $1.658<T<1.705$ (PBC) with $N=10$, and for $N=12$ (PBC) at $T=1.68$ all partial transposes are positive while $\omega_{\cmc}=0.994$. The detection of fully PPT entanglement by the CMC is therefore not a boundary effect. \Cref{fig:N_12} shows the CMC and the PPT test for the alternating cut for $N=12$ at the same temperature. For both boundary conditions, the PPT test detects every point detected by the CMC at this temperature.

The remaining results complement \cref{sec:analytic-witness}: the internal structure of the optimal matrices, the effect of open boundaries and of embedding a subchain in a larger system, and a noncritical point.
\Cref{fig:cmc-optimal-Z}(a) shows, for the two reference points and PBC, the covariance matrix, the optimal matrix $Z^\star$, and the entries of $Z^\star$ as a function of lattice distance.  The optimal matrix selects one Fourier component per spin sector.  Accordingly, its entries have constant modulus $1/(2N)$ as a function of distance, with alternating sign for the components at $q=\pi$, as expected from \cref{eq:mode-projector}.

\Cref{tab:cmc-boundary-values} compares periodic and open chains.  For PBC, \cref{thm:reduction} applies and the optimal value coincides with \cref{eq:exact-cmc}.  For OBC, the covariance matrix is not circulant, the optimal matrix is not exactly a Fourier projector, and the SDP can assign different weights to sites near the edges (\cref{fig:cmc-optimal-Z}(b)).  The optimal values change only moderately, and the three-mode witnesses \cref{eq:three-mode-witness} remain close to optimal.  The boundary-induced inhomogeneity is most visible at the Ising point.

\begin{table*}[t]
\centering
\caption{Optimal CMC values for $N=10$ and $T=0.1$, together with the best three-mode witness \cref{eq:three-mode-witness} with momenta in $\{0,\pi\}$. Values below one certify entanglement.  At the noncritical point the exact optimum for PBC involves an incommensurate momentum (see text).}
\label{tab:cmc-boundary-values}
\begin{tabular}{@{}lcccc@{}}
\toprule
 & \multicolumn{2}{c}{PBC} & \multicolumn{2}{c}{OBC} \\
\cmidrule(lr){2-3}\cmidrule(lr){4-5}
Parameters $(\gamma,h)$ & $\omega_{\cmc}$ & three-mode & $\omega_{\cmc}$ & three-mode \\
\midrule
Ising point $(1,1)$ & $0.583$ & $0.583$ & $0.460$ & $0.474$ \\
XX point $(0,0)$ & $0.154$ & $0.154$ & $0.140$ & $0.144$ \\
Noncritical point $(0.5,0.5)$ & $0.778$ & $0.788$ & $0.772$ & $0.794$ \\
\bottomrule
\end{tabular}
\end{table*}

\begin{figure*}[t]
\centering
\includegraphics[width=\linewidth]{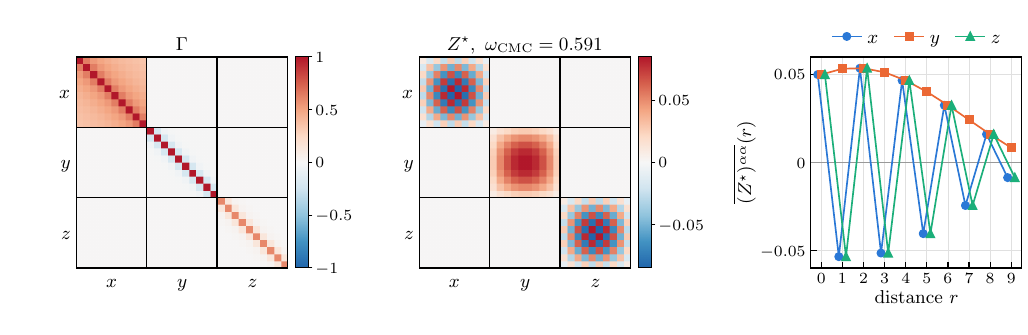}
\caption{Covariance matrix $\Gamma$, optimal matrix $Z^\star$, and entries of the spin blocks $(Z^\star)^{\alpha\alpha}$ averaged over pairs of sites at distance $r$, for a window of $10$ spins in the middle of an open chain of $200$ spins at the Ising point and $T=0.1$.  The approximate translation invariance of the covariance matrix is not inherited by the unrestricted optimal matrix. Layout as in \cref{fig:cmc-optimal-Z}.}
\label{fig:cmc-subchain}
\end{figure*}

A second way to reduce boundary effects is to restrict the covariance matrix to a central window of a much larger open chain.  For a window of $10$ sites in a chain of $200$ sites at the Ising point, the covariance matrix is approximately block-Toeplitz, as expected in the bulk.  The unrestricted optimal matrix does not inherit this approximate symmetry (\cref{fig:cmc-subchain}): it exploits the finite window, and imposing a Toeplitz ansatz could only increase the optimal value.
In the thermodynamic limit, however, the optimal value of such windows approaches the value of the infinite chain, \cref{eq:exact-cmc} with the structure factors \cref{eq:structure-factor-tdl}, since Fourier modes with smooth envelopes become asymptotically optimal.  Conversely, a window of the infinite chain can never do better
and the optimal value of any window is bounded from below by the thermodynamic-limit value.

\begin{figure*}[t]
\centering
\includegraphics[width=\linewidth]{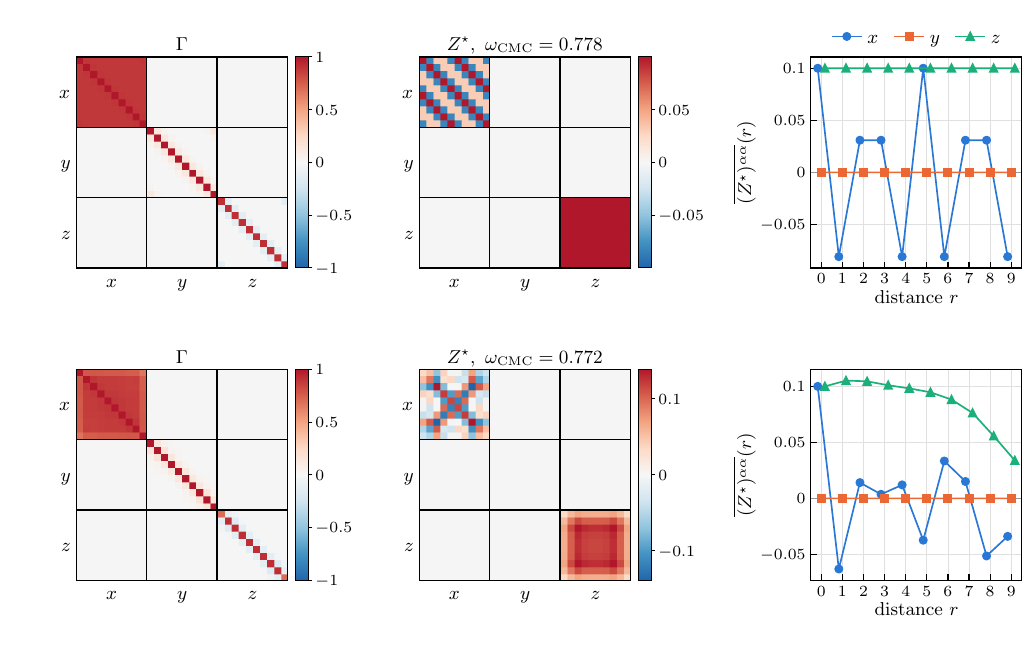}
\caption{Covariance matrix $\Gamma$, optimal matrix $Z^\star$, and entries of the spin blocks $(Z^\star)^{\alpha\alpha}$ averaged over pairs of sites at distance $r$ at the noncritical point $(\gamma,h)=(0.5,0.5)$ for $N=10$ and $T=0.1$, for PBC (upper row) and OBC (lower row).  For PBC, $\omega_{\cmc}=0.778$ is attained by the two-component witness of \cref{thm:reduction} with the incommensurate momentum $q_x=4\pi/5$; for OBC, $\omega_{\cmc}=0.772$. Layout as in \cref{fig:cmc-optimal-Z}.}
\label{fig:cmc-noncritical}
\end{figure*}

To see how the witness changes away from the critical regions, we consider the gapped point $(\gamma,h)=(0.5,0.5)$ inside the ordered region at $T=0.1$. It lies inside the circle $h^2+\gamma^2=1$, where correlations oscillate incommensurately~\cite{KurmannThomasMuller1982, GiampaoloAdessoIlluminati2008}.  For the periodic chain with $N=10$ the three minimal structure factors are $s_x=0.111$ at the incommensurate momentum $q_x=4\pi/5$, $s_y=0.797$ at $q_y=\pi$ and $s_z=0.667$ at $q_z=0$.  Since $s_y>s_x+s_z$, \cref{thm:reduction} gives the two-component witness $\omega_{\cmc}=s_x+s_z=0.778$, with an optimal matrix of eigenvalues $(1,\frac12,\frac12)$: a rank-two block for the incommensurate $x$ mode and a rank-one block for the uniform $z$ mode.  This is exactly what the numerical optimizer returns (\cref{fig:cmc-noncritical}), whereas the best three-mode witness with momenta in $\{0,\pi\}$ gives $0.788$.  The CMC thus still detects low-temperature entanglement, but the violation is appreciably weaker than at the XX point.  At higher temperatures the minima become commensurate again, and already at $T=0.5$ the optimal momenta in the thermodynamic limit are $(\pi,\pi,0)$.

\section{Genuine multipartite entanglement}
\label{app:gme}

In this appendix we collect the material on genuine multipartite entanglement, which we calculate only for small chains.

As mentioned in \cref{subsec:entanglement-framework}, one can design relaxations based on the PPT criterion that also provide a test of {\it genuine multipartite entanglement} (GME). The latter is, as the name suggests, a notion that filters out any form of entanglement that comes just from bipartitions. More precisely, GME is defined as follows: A pure state is biseparable if it is a product $\ket\psi=\ket{\phi_A}\otimes\ket{\chi_{\bar A}}$ with respect to some bipartition $A|\bar A$, and a mixed state is biseparable if it is a convex combination of biseparable pure states, which may be products with respect to different bipartitions~\cite{GuehneToth2009}. A state that is not biseparable is called GME: its entanglement cannot be produced by mixing states in each of which some group of spins is unentangled from the rest. Since a state that is separable across $A|\bar A$ is also PPT across it, every biseparable state is a convex combination $\sum_Ap_A\varrho_A$ of states $\varrho_A$ that are PPT across $A|\bar A$. Hence, states outside the convex hull of states that are PPT across some bipartition, the PPT mixtures, are GME~\cite{JungnitschMoroderGuhne2011}.

In the numerics we use the corresponding witnesses. For GME, a witness $W$ is called fully decomposable if, for every bipartition $A|\bar A$, it can be written as $W=P_A+Q_A^{T_A}$ with $P_A\succeq0$ and $Q_A\succeq0$~\cite{JungnitschMoroderGuhne2011}. For a state $\varrho_A$ that is PPT across $A|\bar A$ one then has $\Tr(W\varrho_A)=\Tr(P_A\varrho_A)+\Tr(Q_A\varrho_A^{T_A})\geq0$, so that fully decomposable witnesses are nonnegative on all PPT mixtures. The optimal one is found from the SDP
\begin{equation}
\begin{split}
v_{\rm PPTmixer}=\min\Bigl\{\Tr(W\varrho):{}&\ \Tr W=1,\ W=P_A+Q_A^{T_A},\\
&\ P_A\succeq0,\ Q_A\succeq0\ \text{for all}\ A\Bigr\},
\end{split}
\label{eq:ppt-mixer-sdp}
\end{equation}
where $A$ runs over all bipartitions $A|\bar A$, and $v_{\rm PPTmixer}<0$ certifies that $\varrho$ is not a PPT mixture and hence GME~\cite{JungnitschMoroderGuhne2011}.

\begin{figure*}[t]
\centering
\includegraphics[width=\linewidth]{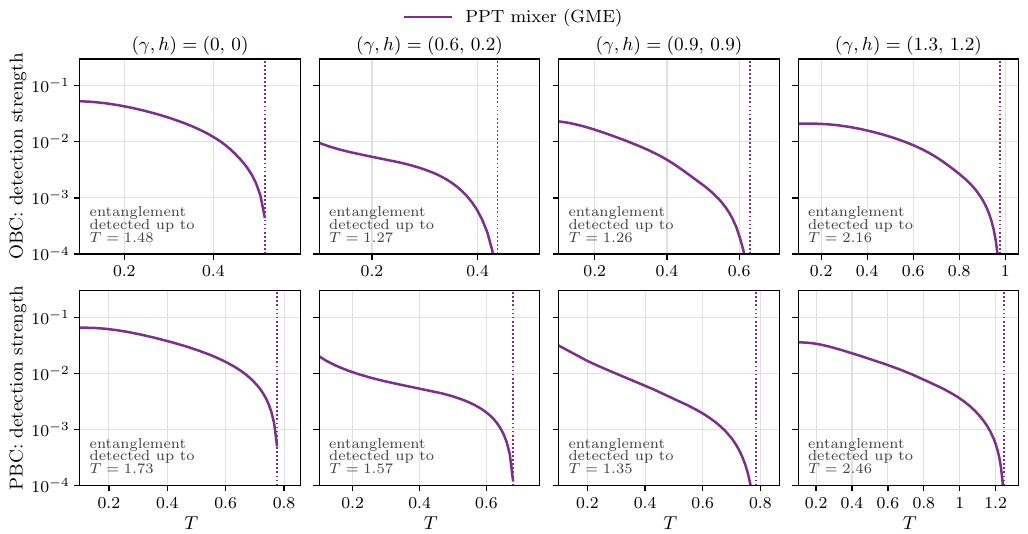}
\caption{Genuine multipartite entanglement certified by the PPT-mixture relaxation \cref{eq:ppt-mixer-sdp} for the thermal states of \cref{fig:criteria-comparison,fig:criteria-comparison-app} with $N=4$, for open (top row) and periodic (bottom row) boundary conditions, at the XX point and at the three points of \cref{fig:criteria-comparison-app}: $\max(0,-v_{\rm PPTmixer})$ on a logarithmic scale. Dotted vertical lines mark the highest temperature at which GME is certified, and each temperature axis ends slightly above it. The numbers in the lower left corners give the highest temperature at which any of the three criteria of \cref{fig:criteria-comparison,fig:criteria-comparison-app} detects entanglement at the same point, which lies outside the plotted range.}
\label{fig:gme}
\end{figure*}

As shown in \cref{fig:gme}, for $N=4$ the PPT mixer certifies GME at all test points at low temperatures, and it does so up to $T\simeq0.4$--$1.0$ for OBC and $T\simeq0.7$--$1.25$ for PBC, i.e., up to between about $35\%$ and $60\%$ of the highest temperature at which entanglement is detected at the same point. That GME is certified in a smaller temperature range is expected, since GME states form a subset of the entangled states. Moreover, a state that is not a PPT mixture is, in particular, not PPT across any bipartition, so that the PPT mixer can never detect more states than the PPT test. In all our phase scans it also detects fewer states than the CMC and the moment hierarchy. We include GME only as a benchmark for the strongest form of multipartite entanglement: the CMC as formulated in \cref{subsec:entanglement-framework}, and hence all witnesses derived in \cref{sec:analytic-witness}, certify that a state is not fully separable. Certifying GME from the covariance matrix would require a separate bound for biseparable states, which we do not pursue here.

Furthermore, a detailed analysis of genuine multipartite entanglement in the reduced three- and four-particle states of the XY chain can be found in Ref.~\cite{HofmannOsterlohGuhne2014}. There, the scaling of $v_{\rm PPTmixer}$ of \cref{eq:ppt-mixer-sdp} close to the quantum phase transitions has also been studied.

\section[\texorpdfstring{Details on the covariance-matrix criterion and proof of \cref{thm:reduction}}{Details on the covariance-matrix criterion and proof of Theorem 1}]{Details on the covariance-matrix criterion and proof of \cref{thm:reduction}}
\label{app:cmc-details}

In this appendix we first fix the normalization of the three-mode witnesses of \cref{prop:three-mode}, then prove \cref{thm:reduction}, and finally show that the covariance matrix is of separable form whenever the optimal witness does not detect entanglement (\cref{cor:certificate}).

For arbitrary momenta $\bm q$, the matrix $\widehat Z=\sum_\alpha P_\alpha$ is positive semidefinite with diagonal blocks $\widehat Z_{jj}=\id_3/N$, and for $\bm q\in\{0,\pi\}^3$ it is a rank-three projector.  Hence $b(\widehat Z)=\sum_j(3/N-1/N)=2$, and the normalized witness is $Z_{\bm q}=\widehat Z/2$, which for $\bm q\in\{0,\pi\}^3$ has nonzero eigenvalues $1/2$ and entries of modulus $1/(2N)$ in the occupied blocks.  This is the normalization of the matrices in \cref{fig:cmc-optimal-Z}, and it reproduces the separable bound in \cref{eq:three-mode-witness}.

We now prove \cref{thm:reduction}.

\begin{proof}[Proof of \cref{thm:reduction}]
As discussed in \cref{subsec:xy-model}, the Gibbs state of the periodic XY chain, and hence its covariance matrix, is invariant under lattice translations, $j\mapsto j+1$ modulo $N$, under the spin parity $P_z$, and under complex conjugation in the $\sigma^z$ basis. The proof has three steps. We first show that the optimal witness can be chosen with the same symmetries as the state, then that such a symmetric witness is characterized by only three weights, one for each spin component, and finally that the resulting bound is attained.

\textit{(i) Symmetrization.}  Translating all sites, or applying the same rotation or reflection of the Bloch sphere at every site, maps a witness $Z$ to a new matrix $Z'=OZO^\T$, where $O$ is the corresponding orthogonal transformation of the $3N$ local Pauli operators. Such a transformation only reshuffles and rotates the diagonal blocks $Z_{jj}$. Hence $Z'$ is again positive semidefinite with $b(Z')=b(Z)$, and if $\Gamma$ is invariant it gives the same value, $\Tr(Z'\Gamma)=\Tr(Z\Gamma)$. Moreover, the admissible witnesses of \cref{eq:cmc-optimum} form a convex set, because $b$ is a concave function (the largest eigenvalue is a convex one). Averaging an optimal witness over all its transformed copies therefore gives a symmetric witness that is still admissible and optimal. The transformations relevant here are the lattice translations, the $\pi$ rotation ${\rm diag}(-1,-1,1)$ (spin parity), and the reflection ${\rm diag}(1,-1,1)$ (complex conjugation), applied to all sites. Invariance under these two transformations removes all couplings between different spin components, so that both matrices are block diagonal, $\Gamma=\bigoplus_\alpha\Gamma^{\alpha\alpha}$ and $Z=\bigoplus_\alpha Z^{\alpha\alpha}$. Invariance under translations makes every block a real symmetric circulant matrix, whose entries depend only on the distance $j-l$. All blocks are therefore diagonalized by the same Fourier modes, with eigenvalues $\sfac_\alpha(q)$ and $\zeta_\alpha(q)\geq0$, respectively.

\textit{(ii) Reduction to three weights.}  In the Fourier basis, the value of a symmetric witness is a sum over momenta. Replacing every structure factor by its minimum $s_\alpha$ can only decrease this sum, so that
\begin{equation}
\begin{gathered}
\Tr(Z\Gamma)=\sum_\alpha\sum_{q\in\mathcal Q_N}\zeta_\alpha(q)\,\sfac_\alpha(q)
\geq\sum_\alpha s_\alpha\kappa_\alpha ,
\\
\kappa_\alpha:=\sum_{q}\zeta_\alpha(q)=N\,Z_{j\alpha,j\alpha}.
\end{gathered}
\end{equation}
Here $\kappa_\alpha$ is the total weight that the witness puts on spin component $\alpha$. The separable bound depends only on these weights, because the diagonal blocks are $Z_{jj}={\rm diag}(\kappa_x,\kappa_y,\kappa_z)/N$ for every $j$. Thus $b(Z)=\sum_\alpha\kappa_\alpha-\max_\alpha\kappa_\alpha$ is the sum of the two smallest weights, and the normalization $b(Z)\geq1$ means that any two weights add up to at least one.  Hence
\begin{equation}
\begin{split}
\!\!\!\!\omega_{\cmc}(\varrho)\geq\min\Bigl\{\sum_\alpha s_\alpha\kappa_\alpha\!\!:\!
\kappa_\alpha\geq0,\kappa_\alpha+\kappa_\beta\geq1\ \!\! \forall \alpha\!\neq\!\beta\!\Bigr\}.
\end{split}
\end{equation}
This is a linear program in three variables. Its feasible region is a polyhedron with the corners $(\frac12,\frac12,\frac12)$, $(1,1,0)$, $(1,0,1)$ and $(0,1,1)$, which extends to infinity only along directions with nonnegative components.  Since $s_\alpha\geq0$, the minimum is attained at one of the corners, which gives the right-hand side of \cref{eq:exact-cmc}.

\textit{(iii) Attainability.}  Finally, the matrix $Z^\star$ of the theorem attains this bound. The matrices $P_\alpha$ are positive semidefinite, have diagonal entries $1/N$ in spin sector $\alpha$, and satisfy $\Tr(P_\alpha\Gamma)=\sfac_\alpha(q_\alpha)=s_\alpha$.  Therefore $b(Z^\star)=\sum_\alpha\kappa_\alpha-\max_\alpha\kappa_\alpha=1$ and $\Tr(Z^\star\Gamma)=\sum_\alpha\kappa_\alpha s_\alpha$ equals the lower bound. In particular, if the first entry of \cref{eq:exact-cmc} is the smallest, then $Z^\star=Z_{\bm q}$ with the minimizing momenta $\bm q=(q_x,q_y,q_z)$ and, by \cref{eq:momentum-spin}, $\Tr(Z^\star\Gamma)=\omega_{\bm q}(\varrho)$. The separable bound in \cref{eq:three-mode-witness} is \cref{eq:cmc-dual-bound} for $Z=Z_{\bm q}$, since $b(Z_{\bm q})=1$, see above.
\end{proof}

\begin{corollary}[Explicit separable covariance matrix]
\label{cor:certificate}
Under the assumptions of \cref{thm:reduction}, let $\omega_{\cmc}(\varrho)\geq1$. Then there are numbers $x_\alpha\geq\max\{0,1-s_\alpha\}$ with $x_x+x_y+x_z=1$, and with $X={\rm diag}(x_x,x_y,x_z)$ one has
\begin{equation}
\begin{gathered}
\Gamma(\varrho)=\bigoplus_{j=1}^N(\id_3-X)+K,
\\
K=\bigoplus_{\alpha}\left[\Gamma^{\alpha\alpha}-(1-x_\alpha)\id_N\right]\succeq0 .
\end{gathered}
\label{eq:separable-cm}
\end{equation}
Hence $\Gamma(\varrho)$ is a separable covariance matrix in the sense of \cref{eq:cmc-primal}, and no CMC witness, i.e.\ no local uncertainty relation of the form \cref{eq:cmc-lur} with the bound \cref{eq:cmc-dual-bound}, can detect entanglement in $\varrho$.
\end{corollary}

\begin{proof}
The numbers $x_\alpha$ exist if and only if $\sum_\alpha\max\{0,1-s_\alpha\}\leq1$.  If all $s_\alpha\leq1$, this reads $s_x+s_y+s_z\geq2$. If exactly one of them exceeds one, it reads $s_\alpha+s_\beta\geq1$ for the other two components, and if two or more exceed one, it holds trivially.  In all cases the condition is equivalent to the right-hand side of \cref{eq:exact-cmc} being at least one.  Moreover, $\Gamma^{\alpha\alpha}\succeq s_\alpha\id_N$ implies $\Gamma^{\alpha\alpha}-(1-x_\alpha)\id_N\succeq (s_\alpha-1+x_\alpha)\id_N\succeq0$.
\end{proof}

\section{Details of the quench calculation}
\label{app:quench}

In this appendix we give the details of the quench of \cref{sec:quench}. The initial state $\varrho(0)$ is the Gibbs state \eqref{eq:thermal-state} at temperature $T_i$ of $H_i$, i.e., of the Hamiltonian \eqref{eq:xy-hamiltonian} with $\gamma=1$ and transverse field $h_i$, and it evolves unitarily under $H_f$, which has the same form with the field $h_f$. Both Hamiltonians are quadratic in the Majorana operators defined below, so that $\varrho(t)$ remains a fermionic Gaussian state.
Setting $a_{2j-1}=(\prod_{l<j}\sigma^z_l)\,\sigma^x_j$ and $a_{2j}=(\prod_{l<j}\sigma^z_l)\,\sigma^y_j$, with the Jordan--Wigner strings of \cref{eq:jw}, we can write the Hamiltonian as $H=\tfrac{\ii}4\,a^{\T}\mathcal Aa$ with the real antisymmetric matrix $\mathcal A$ whose nonzero entries above the diagonal are
\[
\mathcal A_{2j-1,2j}=2h,\qquad \mathcal A_{2j,2j+1}=1+\gamma,\qquad \mathcal A_{2j-1,2j+2}=-(1-\gamma).
\]
We denote by $\mathcal A_i$ and $\mathcal A_f$ the matrices $\mathcal A$ of $H_i$ and $H_f$, i.e., with $h=h_i$ and $h=h_f$, respectively.
Gaussian states are fixed by $M=\tfrac{\ii}2\mean{[a,a^{\T}]}$, with
\[
M_\beta=\ii\tanh\!\big(\tfrac\beta2\, \ii\mathcal A\big),\qquad M(t)=\e^{\mathcal A_ft}M(0)\,\e^{\mathcal A_f^{\T}t}.
\]
The steady state is obtained from $M(0)$ by removing all coherences between eigenvectors of $\ii\mathcal A_f$ with different eigenvalues, which yields the GGE of the free-fermion chain~\cite{Rigol2007,CEF2011,EsslerFagotti2016}. The energy $\mean{H_f}=-\Tr(\mathcal A_fM)/4$ is conserved, and it fixes the temperature $T_{\rm eff}$ of the Gibbs state of $H_f$ with the same energy. In the infinite chain, which is the case we consider here, these relations hold per momentum for $2\times2$ symbols. Any window of $L$ consecutive spins is again Gaussian, and a detected window certifies entanglement of the infinite chain.

Spin correlators are Pfaffians of submatrices of $M$~\cite{Wimmer2012}. The Pfaffian of an antisymmetric $2n\times2n$ matrix $B$ is $\operatorname{Pf}(B)=\frac{1}{2^nn!}\sum_{\pi}\operatorname{sgn}(\pi)\prod_{k=1}^{n}B_{\pi(2k-1)\pi(2k)}$, where the sum runs over all permutations $\pi$ of $\{1,\dots,2n\}$, and it satisfies $\operatorname{Pf}(B)^2=\det B$. By Wick's theorem, the expectation value of a product of $2n$ distinct Majorana operators is $\mean{a_{m_1}\cdots a_{m_{2n}}}=(-\ii)^n\operatorname{Pf}(M_{\bm m})$, where $M_{\bm m}$ is the submatrix of $M$ with rows and columns $m_1,\dots,m_{2n}$. Through the Jordan--Wigner strings, spin correlators become expectation values of this form, e.g.
\begin{gather*}
\sigma^{\alpha}_i\sigma^{\beta}_j=(-\ii)^{j-i}a_{\mu_{\alpha}(i)}\,(a_{2i-1}\cdots a_{2j-2})\,a_{\mu_{\beta}(j)},\\
\mu_x(i)=2i-1,\quad \mu_y(i)=2i.
\end{gather*}
For thermal (real) states the Pfaffians reduce to the determinant formulas of \cref{app:xy-solution}. After a quench, we have $\varrho^*(t)=\varrho(-t)$ and thus
the quench creates $x$--$y$ correlations $\mean{\sigma^x\sigma^y}$, so the general Pfaffian form is needed.
In particular, this also makes the optimal CMC witness more complicated, even though the structure-factor witnesses of \cref{eq:three-mode-witness} remain valid upper bounds on $\omega_{\cmc}$.

The optimum itself can be obtained numerically, with an SDP that can be simplified taking into account translation invariance. Averaging any feasible $Z$ over translations preserves feasibility and leaves the objective unchanged, so the optimal $Z$ can be taken block diagonal in momentum. This gives an SDP over $3\times3$ Hermitian blocks $\hat Z(q)\succeq0$,
\begin{equation}
\begin{split}
\omega_{\cmc}=\min\Bigl\{&\frac1{n_q}\sum_q\Tr\bigl[\hat Z(q)\hat\Gamma(q)\bigr]:\\
&\hat Z(q)\succeq0,\ \Tr Z_0-\lambda_{\max}(Z_0)\geq1\Bigr\},
\end{split}
\label{eq:app-block-sdp}
\end{equation}
where $\hat\Gamma(q)$ is the $3\times3$ Fourier transform of the covariance matrix, $Z_0=\frac1{n_q}\sum_q\hat Z(q)$ the diagonal block of $Z$ at each site, and $q$ runs over $n_q$ equally spaced momenta.

\begin{figure*}[t]
\centering\includegraphics[width=1.0\linewidth]{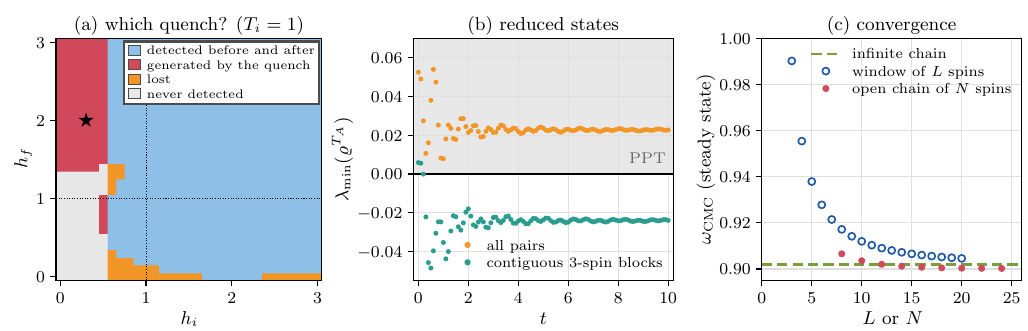}
\caption{(a)~Classification of Ising quenches at $T_i=1$ by the steady state; the star marks the example of \cref{fig:quench}. (b)~Smallest partial-transpose eigenvalue of all two-spin and contiguous three-spin reduced states (grey: PPT). (c)~Steady-state $\omega_{\cmc}$ for windows of the infinite chain and for open chains.}
\label{fig:quench-app}
\end{figure*}

Since $\Gamma$ is concave in the state, the steady-state value bounds the time average of $\omega_{\cmc}(t)$ from above.
We summarize our numerical results in \cref{fig:quench-app}. At $T_i=1$, where the initial state is not detected for $h_i<0.51$, entanglement is generated for $h_i\lesssim0.5$ and $h_f\gtrsim1.4$.
Stronger quenches give a more strongly entangled steady state (and at the same time inject more energy). No pair of spins is ever entangled, but contiguous three-spin blocks are entangled from $t\approx0.2$ on. Windows of $L$ spins converge to the infinite-chain steady state from above, and $L=3$ already certifies it. Open chains converge to $\omega_{\cmc}\simeq0.900$ by $N\approx16$.

For an open chain of $N = 8$ spins, starting from a state which is fully PPT across all bipartitions, we obtain a post-quench state that is NPT across some bipartition at each of the $40$ times $t=0.5,1,\dots,20$. The CMC also detects those states except $t=4.5$ and $5$.

\section{Best separable approximation and geometric measure}
\label{app:quantification-details}
\label{app:bsa-dual}
\label{app:bsa-geometric-roofs}
\label{app:geometric-legendre-details}

Writing the separable part as an unnormalized separable operator $\widetilde\varrho_{\rm sep}$, the separable weight of the BSA is the conic program
\begin{equation}
\begin{split}
\lambda_{\bsa}(\varrho)&=1-E_{\bsa}(\varrho)\\
&=\max\bigl\{\Tr\widetilde\varrho_{\rm sep}:\ \widetilde\varrho_{\rm sep}\in{\rm cone}(\SEP),\\
&\hphantom{{}=\max\bigl\{}\ \varrho-\widetilde\varrho_{\rm sep}\succeq0\bigr\}.
\end{split}
\label{eq:app-bsa-conic-primal}
\end{equation}
Its dual is
\begin{equation}
\begin{split}
\lambda_{\bsa}(\varrho)=\min\bigl\{\Tr(Y\varrho):{}&\ Y\succeq0,\\
&\ \Tr(Y\varrho_{\rm sep})\geq1\ \forall\varrho_{\rm sep}\in\SEP\bigr\},
\end{split}
\label{eq:app-bsa-conic-dual}
\end{equation}
and the substitution $W=Y-\id$ gives \cref{eq:bsa-witness-dual}.  Any feasible $W$ thus bounds $E_{\bsa}$ from below, and any feasible $\widetilde\varrho_{\rm sep}$ bounds it from above.

The linearized covariance witness used in the proof of \cref{prop:bsa-bound} is a feasible point of \cref{eq:bsa-witness-dual}.  Once its centers $a_\nu$ have been fixed from the data, it is a single linear witness, so no nonlinearity enters the bound.  It is the site-resolved counterpart of the construction for collective spin-squeezing inequalities~\cite{FadelEtAl2021,MatheEtAl2026}.

We now derive the convex-roof representations \cref{eq:convex-roofs}. Let $e(\psi)=0$ for product states and $e(\psi)=1$ otherwise.  Given any pure-state decomposition $\varrho=\sum_kp_k\ketbra{\psi_k}$, the product members form a separable part of weight $1-t$, with $t$ the total weight of the entangled members, so $E_{\bsa}(\varrho)\leq\sum_kp_ke(\psi_k)$.  Conversely, refining the two parts of any decomposition $\varrho=(1-t)\varrho_{\rm sep}+t\tau$ into pure product states and pure states, respectively, gives an ensemble with $\sum_kp_ke(\psi_k)\leq t$.  Hence $E_{\bsa}$ is the convex roof of $e$, as stated in \cref{eq:convex-roofs}.  The same refinement shows that minimizing the weight of the entangled part after multiplying each of its pure components by $E_G(\psi)=1-\Lambda^2(\psi)$ gives exactly the convex-roof geometric measure.  Since $0\leq E_G(\psi)\leq e(\psi)$, the ordering \cref{eq:geometric-bsa-order} follows.

The lower bound \cref{eq:geometric-legendre-bound} on $E_G$ follows from the Legendre transform. For a convex function $E$ on states, the Fenchel inequality $E(\varrho)\geq\Tr(X\varrho)-\widehat E(X)$ holds with $\widehat E(X)=\sup_\varpi[\Tr(X\varpi)-E(\varpi)]$.  For a convex-roof measure the supremum may be restricted to pure states, and for the geometric measure
\begin{equation}
\begin{split}
\widehat E_G(X)&=\sup_{\ket\psi}\Bigl[\mean{X}_\psi-1+\max_{\ket\phi\ {\rm product}}
|\braket{\phi}{\psi}|^2\Bigr]\\
&=-1+\max_{\ket\phi\ {\rm product}}\lambda_{\max}\bigl(X+\ketbra\phi\bigr).
\end{split}
\label{eq:app-geometric-conjugate}
\end{equation}
Choosing $X=\lambda W$ and optimizing over $\lambda$ gives \cref{eq:geometric-legendre-bound}~\cite{GuhneReimpellWerner2007, GuhneReimpellWerner2008}.  The choice $\lambda=0$ shows that the bound is nonnegative.  If several expectation values are known, one may use a separate multiplier for each of them, which can only improve the bound.

\begin{table*}[t]
\centering
\caption{Lower bounds on the geometric measure $E_G$ from \cref{eq:geometric-legendre-bound} for the optimal witness values $\omega_{\cmc}$ of the periodic chain with $N=10$ at $T=0.1$ (\cref{tab:structure-factor-minima}). The inner maximum is evaluated for chains of $N$ spins, either at the Néel-type states suggested by the numerical search or with the PPT relaxation, which gives a certified bound.}
\label{tab:geometric-bounds}
\begin{tabular}{@{}lcccc@{}}
\toprule
 & $\omega_{\cmc}$ & $N$ & Néel-type states & PPT relaxation \\
\midrule
XX point $(0,0)$ & $0.154$ & $6$ & $0.3575$ & $0.3575$ \\
 & $0.154$ & $8$ & $0.3834$ & $0.3833$ \\
 & $0.154$ & $10$ & $0.4009$ & $0.3993$ \\
Ising point $(1,1)$ & $0.583$ & $10$ & $0.0620$ & $0.0302$ \\
\bottomrule
\end{tabular}
\end{table*}

Let us consider the witness $W=\frac2N\sum_\alpha J^2_\alpha(q_\alpha)-\id$, where $q_\alpha\in \{0, \pi\}$, which corresponds to the case where the first moments vanish, as it is for our reference Ising and XX points.
At the XX point, the local unitary $U=\prod_{j\ {\rm even}}\sigma_j^z$, i.e., a $\pi$ rotation about the $z$ axis of every second spin, maps $J_x(\pi)$, $J_y(\pi)$ and $J_z(0)$ to $-J_x(0)$, $-J_y(0)$ and $J_z(0)$, respectively. At the Ising point, $U=\prod_{j\ {\rm even}}\sigma_j^y$, i.e., a $\pi$ rotation about the $y$ axis of every second spin, maps $J_x(\pi)$, $J_y(0)$ and $J_z(\pi)$ to $-J_x(0)$, $J_y(0)$ and $-J_z(0)$. In both cases, the staggered rotation $U$, which leaves the set of product states invariant, turns the witness into a function of the total spin $\bm J=(J_x(0),J_y(0),J_z(0))$,
\begin{equation}
\begin{gathered}
W'=UWU^\dagger=\frac2N\bm J^2-\id=\sum_J w_J\,\Pi_J ,
\\
w_J=\frac{2J(J+1)}N-1,
\end{gathered}
\label{eq:app-total-spin-witness}
\end{equation}
where $\Pi_J$ projects onto the subspace with total spin $J$, i.e., onto the eigenspace of $\bm J^2$ with eigenvalue $J(J+1)$.  For $\lambda\geq0$ the fully polarized product state attains the operator-norm bound $1+\lambda w_{N/2}$ on the inner maximum, so that $\lambda\geq0$ never improves on zero, and it suffices to consider $\lambda<0$.

A numerical maximization of the inner problem over product states, with random initial states, for $N=4$, $6$, $8$ and $10$ and a range of $\lambda<0$ suggests that the maximum is attained at Néel-type states, with $N/2$ spins along a unit vector $\bm n$ and $N/2$ spins along $-\bm n$. A rigorous lower bound, however, requires an upper bound on the inner maximum. Such a bound can be obtained by relaxing the set of product states to the set of states that are PPT across every bipartition. Since $W'$ is invariant under permutations and collective rotations, this relaxation can be addressed via an SDP, which reduces to a small linear program over the weights of the total-spin sectors, and it gives values similar to those of the numerical search (\cref{tab:geometric-bounds}).

For the witness value of the periodic chain with $N=10$ at the XX point and $T=0.1$ (\cref{tab:structure-factor-minima}), and chains of $N=6$, $8$ and $10$ spins, the certified bound nearly coincides with the one obtained from the Néel-type states, so that the relaxation is essentially tight in this regime, where the optimal multiplier $|\lambda|$ is large.  At the Ising point, where $\omega_{\cmc}$ is closer to one and the optimal multiplier is smaller, the relaxation is looser, but it still certifies a nonzero geometric measure.

Upper bounds on $E_{\bsa}$ and $E_G$ follow from explicit separable states. Product states can be symmetrized with the symmetry group of $\varrho_\beta$ (translations, parity, lattice reflection, complex conjugation and, on the XX line, the $U(1)$ rotations), which preserves separability and loses nothing at the optimum~\cite{VollbrechtWerner2001,StreltsovKampermannBruss2010}.  The weights of a growing dictionary of such orbits can be optimized by SDP, either for the separable weight in \cref{eq:app-bsa-conic-primal} or for the fidelity with $\varrho$, and new product states can be generated from the dual solution by column generation~\cite{KarnasLewenstein2001,OhstEtAl2024,MatheEtAl2026}.  The witness suggests good starting points.  If both the witness and the decomposition were optimal, complementary slackness would require every product state in the separable part to satisfy $\bra vW\ket v=0$, i.e.\ to have zero total Bloch vector in the rotated frame, and the remainder $\varrho-(1-E_{\bsa})\varrho_{\rm sep}$ to be supported in the kernel of $W+\id$, i.e.\ in the singlet sector $J=0$.

\section*{Data and code availability}

The numerical data and the code used to generate the detection masks, the optimal covariance witnesses, and the structure factors will be deposited in a public repository together with the final version of the manuscript.

{\small
\bibliographystyle{SciPost_bibstyle}
\bibliography{biblio}
}

\end{document}